\documentclass[%
 reprint,
superscriptaddress,
showpacs,%preprintnumbers,
nofootinbib,
 amsmath,amssymb,aps,
floatfix,
]{revtex4-1}

\usepackage[dvipsnames]{xcolor}
\usepackage{soul}% Contains the strikeout command \st
\usepackage{graphicx}% Include figure files
\usepackage{dcolumn}% Align table columns on decimal point
\usepackage{bm}% bold math
\usepackage{verbatim}
\usepackage[caption=false]{subfig}

\usepackage[colorlinks=true]{hyperref}% add hypertext capabilities
\hypersetup{
    colorlinks,
    linkcolor=blue!50!black,          % color of internal links (change box color with linkbordercolor)
    citecolor=blue!50!black,        % color of links to bibliography
    urlcolor=blue!50!black           % color of external links
}
\usepackage{tabu}
\usepackage{multirow}
\usepackage{braket}

\graphicspath{{NSimages/}}

\begin{document}

%\preprint{APS/123-QED}

\title{Eigenvalues and eigenstates of the many-body collective neutrino oscillation problem}% Force line breaks with \\

\author{Amol V. Patwardhan}
 \email{apatwardhan2@berkeley.edu}
\affiliation{Department of Physics, University of California, Berkeley, California 94720-7300, USA}%Lines break automatically or can be forced %with \\
\affiliation{%
Department of Physics, University of Wisconsin, Madison, Wisconsin 53706, USA
}%

\author{Michael J. Cervia}
 \email{cervia@wisc.edu}
\affiliation{%
Department of Physics, University of Wisconsin, Madison, Wisconsin 53706, USA
}%
% \altaffiliation[Also at ]{Physics Department, XYZ University.}%Lines break automatically or can be forced %with \\

\author{A. Baha Balantekin}%
 \email{baha@physics.wisc.edu}
\affiliation{%
Department of Physics, University of Wisconsin, Madison, Wisconsin 53706, USA
}%


%\affiliation{
%\affiliation{
%\affiliation{%


\date{\today}% It is always \today, today,

\begin{abstract}

We demonstrate a method to systematically obtain eigenvalues and eigenstates of a many-body Hamiltonian describing collective neutrino oscillations. The method is derived from the Richardson-Gaudin framework, which involves casting the eigenproblem as a set of coupled nonlinear \lq\lq Bethe Ansatz equations\rq\rq, the solutions of which can then be used to parametrize the eigenvalues and eigenvectors. The specific approach outlined in this paper consists of defining auxiliary variables that are related to the Bethe-Ansatz parameters, thereby transforming the Bethe-Ansatz equations into a different set of equations that are numerically better behaved and more tractable. We show that it is possible to express not only the eigenvalues, but also the eigenstates, directly in terms of these auxiliary variables without involving the Bethe Ansatz parameters themselves. %, and subsequently outline a procedure to obtain numerical solutions to these equations. 
In this paper, we limit ourselves to a two-flavor, single-angle neutrino system.
%\begin{description}
%\end{description}
\end{abstract}
\maketitle

%\tableofcontents

\section{Introduction} \label{sec:intro}

Experimental evidence has ascertained that neutrinos can undergo flavor oscillations, as a result of mass-differences between propagation eigenstates, which are distinct from the eigenstates of the weak interaction~\cite{Fukuda:1998mi,Ahmad:2001an,An:2012eh}. Additionally, it has been shown that the mixing between the two sets of eigenstates can be modified in the presence of coherent forward scattering of neutrinos on charged leptons~\cite{Wolfenstein78, Mikheyev85, Haxton:1986uq, Bethe:1986ej}, as well as coherent neutrino-neutrino forward scattering~\cite{Fuller87, Notzold:1988fv, Pantaleone:1992eq, Sigl:1993fr, Raffelt:1993kx, Samuel:1993sf}. The latter contribution is of particular interest on account of its nonlinear nature, and becomes relevant in environments with high neutrino fluxes, such as the hot and dense early universe~\cite{Kostelecky:1993ys, Kostelecky:1995rz, Dolgov:2002ab, Ho:2005vj, McKellar:1994uq, Enqvist:1991yq, Pastor:2002zl, Abazajian:2002lr, Hannestad:2006qd, Johns:2016enc}, as well as in compact-object systems, such as the neutrino emission accompanying a core-collapse supernova explosion or black hole-neutron star or binary neutron star merger. The interplay between the linear and nonlinear terms can result in various forms of interesting collective flavor oscillation phenomena~\cite{Duan:2009cd, Duan:2010fr, Chakraborty:2016a, Balantekin:2018mpq}, including spectral splits/swaps~\cite{Duan06a, Duan06b, Duan06c, Duan07a, Duan07b, Duan07c, Duan:2008qy, Raffelt07, Raffelt:2007, Fogli07, Dasgupta:2007ws, Dasgupta:2008qy, Dasgupta09, Friedland:2010yq, Dasgupta:2010cd, Galais:2011gh, Mirizzi:2013rla, Mirizzi:2013wda, Pehlivan:2016lxx, Tian:2017xbr, Das:2017iuj, Birol:2018qhx, Cirigliano:2018rst}, matter-neutrino resonances~\cite{Malkus:2012, Malkus:2014, Wu:2016, Stapleford:2016jgz, Malkus:2016, Frensel:2016fge, Vaananen:2016, Zhu:2016mwa, Johns:2016enc, Shalgar:2017pzd, Vlasenko:2018irq}, and fast flavor oscillations arising from spatial or temporal instabilities~\cite{Sawyer:2005yg, Chakraborty:2016lct, Capozzi:2016oyk, Wu:2017qpc, Sen:2017ogt, Dighe:2017sur, Abbar:2017pkh, Wu:2017drk, Dasgupta:2017oko, Dasgupta:2018ulw, Capozzi:2018clo, Abbar:2018beu, Shalgar:2019kzy}.

Collective oscillations of neutrinos have been investigated in literature predominantly using the \lq\lq mean field\rq\rq\ approach, wherein each neutrino is considered to be interacting with a background mean field composed of all other neutrinos. However, it has been pointed out that this problem also lends itself to a full many-body description~\cite{Bell:2003mg, Friedland:2003dv, Friedland:2003eh, Balantekin:2007kx, Pehlivan:2011hp, Birol:2018qhx, Cervia:2019nzy, Balantekin:2018mpq}. Such a description is much more complete than the mean field approach, in the sense that it operates within a larger Hilbert space, and captures exclusive many-body effects such as the formation and the evolution of entangled neutrino states. The many-body Hamiltonian describing neutrino oscillations exhibits a mathematical analogy to that of a one-dimensional spin chain with one-particle and \lq\lq long-range\rq\rq\ two-particle interactions (in momentum space), as well as the separable pairing Hamiltonians describing nucleon pairs present in different Shell Model orbitals \cite{Pan:1997rw,Balantekin:2007vs}. Eigenvalues and eigenvectors of the latter class of Hamiltonians were already constructed in the 1960s by Richardson \cite{Richardson:1966zza}. This solution was cast into an algebraic form by Gaudin \cite{Gaudin:1976sv}, whose formalism we shall use extensively in this article. 

The article is organized as follows. In Sec.~\ref{sec:hamilt}, we present the many-body neutrino oscillation Hamiltonian with vacuum and self-interaction terms, and simplify it using the single-angle approximation. Sec.~\ref{sec:bethe} introduces the Bethe Ansatz method, also known as the Richardson-Gaudin diagonalization technique, and demonstrate how the eigenvalues and eigenvectors of the Hamiltonian may be systematically expressed using this method. In Sec.~\ref{sec:Lambdaeq}, we re-cast the Bethe Ansatz equations into a different set of simpler, more tractable equations, and also outline a procedure for calculating the eigenvalues and eigenstates of the Hamiltonian in terms of the solutions to these simpler equations. In Sec.~\ref{sec:bethesol}, we describe the analytic and numerical solutions for specific cases. We conclude in Sec.~\ref{sec:concl}. Many of the mathematical details and derivations are presented in the Appendices. 

\section{The many-body neutrino Hamiltonian}~\label{sec:hamilt}

In scenarios where the neutrino scattering rates are sufficiently low for the neutrinos to be essentially free-streaming across the relevant physical scales, the flavor evolution of neutrinos is dominated by coherent forward-scattering processes. In that case, the interacting neutrinos may be described as a many-body Hamiltonian system. Generally speaking, such a Hamiltonian shall consist of terms that represent neutrino oscillations in vacuum, as well as interactions of neutrinos with ordinary background matter and with other neutrinos, along with the corresponding terms for anti-neutrinos. 

In this article, for ease of discussion and to reduce numerical complexity, we make a number of simplifying assumptions. First and foremost, we restrict the discussion to two flavor/mass states of neutrinos, rather than the full three-flavor picture. Additionally, we pick a system consisting only of neutrinos, and no antineutrinos. Furthermore, keeping the environments where neutrino-neutrino interactions are dominant in mind, we ignore the interactions between neutrinos and ordinary (non-neutrino) background matter. With these assumptions, the Hamiltonian may be written as a sum of vacuum and self-interaction terms, $H_\text{vac}$ and $H_{\nu\nu}$, given by
\begin{equation} \label{eq:ham}
H = \sum_{\mathbf{p}}\omega_\mathbf{p} \vec{B}\cdot\vec{J}_{\mathbf{p}} + \frac{\sqrt{2}G_{F}}{V}\sum_{\mathbf{p},\mathbf{q}}\left(1-\cos\vartheta_{\mathbf{p}\mathbf{q}}\right)\vec{J}_{\mathbf{ p}}\cdot\vec{J}_{\mathbf{q}}~,
\end{equation}
where $\vec{B}=(0,0,-1)_{\mbox{\tiny mass}}=(\sin2\theta,0,-\cos2\theta)_{\mbox{\tiny flavor}}$ is a unit vector indicating the direction of the mass basis in isospin space, with $\theta$ being the vacuum mixing angle, and $\omega_\mathbf{p} = {\delta m^2}/{(2|\mathbf{p}|)}$ are the vacuum oscillation frequencies. Here, $\vartheta_{\mathbf{p}\mathbf{q}}$ is the intersection angle between the trajectories of neutrinos with 3-momenta $\mathbf{p}$ and $\mathbf{q}$, $V$ is the quantization volume, and $G_F$ is the Fermi coupling constant. Here, we have defined the neutrino mass-basis isospin operators $\vec{J}_{\mathbf{p}}$ in terms of the Fermionic creation and annihilation 
operators \cite{Balantekin:2007kx}
\begin{gather}
	{J}_{\mathbf{p}}^{+}= a_{1}^{\dagger}(\mathbf{p})a_{2}(\mathbf{p})~, \\
	{J}_{\mathbf{p}}^{-}= a_{2}^{\dagger}(\mathbf{p})a_{1}(\mathbf{p})~, \\
	{J}_{\mathbf{p}}^z=\frac{1}{2}\left(a_{1}^{\dagger}(\mathbf{p})a_{1}(\mathbf{p})-a_{2}^{\dagger}(\mathbf{p})a_{2}(\mathbf{p})\right)~. 
%			\label{Flavor Isospin Operators} 
\end{gather} 

An analogous set of weak isospin operators may be defined in the flavor basis, using the corresponding creation/annihilation operators, which are related to their mass basis counterparts via the following unitary transformation

\begin{eqnarray} 
%			\label{Flavor to Mass Transformation} 
	a_{e}(\mathbf{p}) & = &  \cos\theta\: a_{1}(\mathbf{p})+\sin\theta\: a_{2}(\mathbf{p})\\ 
	a_{x}(\mathbf{p}) & = & -\sin\theta\: a_{1}(\mathbf{p})+\cos\theta\: a_{2}(\mathbf{p}).
\end{eqnarray}

The isospin operators form an SU(2) algebra, obeying the usual commutation relations
\begin{equation} 
	[{J}_{\mathbf{p}}^{+},{J}_{\mathbf{q}}^{-}]=2\delta_{\mathbf{p}\mathbf{q}}{J}_{\mathbf{p}}^z~,\qquad 
	[{J}_{\mathbf{p}}^z,{J}_{\mathbf{q}}^{\pm}]=\pm\delta_{\mathbf{p}\mathbf{q}}{J}_{\mathbf{p}}^{\pm}. 
%			\label{Flavor Isospin Algebra} 
\end{equation} 

At this point, it is important to note that the strength of the interaction between any two neutrinos is dependent on the intersection angle between their momenta, as can seen from the second term of the Hamiltonian in Eq.~(\ref{eq:ham}). This geometric dependence makes the collective neutrino oscillation problem extremely complex, even in the mean field limit. To avoid these complexities, and to facilitate a qualitative understanding of various collective flavor evolution phenomena, the so-called \lq\lq single-angle\rq\rq\ approximation has been frequently adopted in the literature, wherein the angle-dependent interaction is instead replaced by an overall, angle-averaged coupling strength. The implication of this approximation is that it removes any trajectory-dependence from the flavor evolution of neutrinos. It has been demonstrated that the single-angle calculations are able to capture many of the qualitative behaviors observed in the more sophisticated multi-angle calculations. To this end, we define a direction-independent weak isospin operator $\vec{J}_\omega$ as follows:
\begin{equation}
\vec{J}_\omega = \sum_{|\mathbf{p}| = \frac{\delta m^2}{2\omega}} \vec J_{\mathbf{p}}.
\end{equation}

The self-interaction term may then be approximated as
\begin{equation} 
\begin{split}
%			\label{Self Interactions} 
{H}_{\nu\nu} 
%&=\frac{\sqrt{2}G_{F}}{V}\sum_{\mathbf{p},\mathbf{q}}\left(1-\cos\vartheta_{\mathbf{p}\mathbf{q}}\right)\vec{J}_{\mathbf{ p}}\cdot\vec{J}_{\mathbf{q}} \\
&\approx \frac{\sqrt{2}G_{F}}{V} \langle 1-\cos\vartheta_{\mathbf{p}\mathbf{q}} \rangle \, \vec{J}\cdot\vec{J} \\ %\sum_{\hat{\mathbf{p}} \neq \hat{\mathbf{q}}} \vec{J}_{\mathbf{ p}}\cdot\vec{J}_{\mathbf{q}} \\
&\equiv \mu(r) \, \vec{J}\cdot\vec{J},
\end{split}
\end{equation} 
where we have defined $\vec{J} = \sum_\omega J_\omega = \sum_\mathbf{p} J_\mathbf{p}$. The radial dependence of the coupling strength $\mu$ arises due to the geometric dilution of the neutrino fluxes and the narrowing of the intersection angles as one moves further from the source. For example in a \lq\lq neutrino bulb\rq\rq\ model, where neutrinos are assumed to be emitted isotropically from a single spherical emission surface, the dependence is given by
\begin{equation}
\mu(r) \propto \frac12\left( 1 - \sqrt{1-\frac{R_\nu^2}{r^2}}\right)^2,
\end{equation}
where $R_\nu$ is the radius of the neutrinosphere.

\section{Diagonalization and the Bethe Ansatz equations}~\label{sec:bethe}

In the single-angle picture, the Hamiltonian from Eq.~(\ref{eq:ham}) may be re-written as
\begin{equation} \label{eq:saham}
H = \sum_{p}\omega_p\vec{B}\cdot\vec{J}_{p} +\mu(r) \vec{J}\cdot\vec{J},
\end{equation}
where $p$ is now just an index denoting the oscillation frequencies present in the system. It has been shown that this particular Hamiltonian is amenable to diagonalization using the Richardson-Gaudin procedure. To begin with, one may observe that, among the common eigenstates $\ket{j,m}$ of the total weak isospin operators  $\vec{J}^2$ and $J^z$, some can be written as direct products of eigenstates $\ket{j_p,\pm j_p}$ of $\vec{J}_p^2$ and $J_p^z$. For instance, the states
\begin{equation}
\begin{aligned}
\ket{j,+j} &\equiv \bigotimes_p \ket{j_p,+j_p} \\
\ket{j,-j} &\equiv \bigotimes_p \ket{j_p,-j_p},
\end{aligned}
\end{equation}
where $j = \sum_p j_p$, can be shown to be simultaneous eigenstates of $\vec{J}^2$ and each $J_p^z$, and are therefore eigenstates of the Hamiltonian\footnote{Note that, for $j < N/2$, not every state of the form $\ket{j,\pm j}$ is an eigenstate. $j = \sum_p j_p$ is a necessary condition.}. In particular, the highest and lowest weight states $\ket{\frac N2,\frac N2}$ and $\ket{\frac N2,-\frac N2}$, where $N$ is the total number of neutrinos in the system, are eigenstates. These represent the states where all the neutrinos are either isospin-up or isospin-down, i.e., $\ket{\nu_1,\ldots,\nu_1}$ and $\ket{\nu_2,\ldots,\nu_2}$, respectively. The corresponding eigenvalues for these two states can be easily shown to be
\begin{equation}
E_{\pm N/2}= \mp \sum_{p} \omega_p \, \frac{N_p}{2}+\mu \, \frac N2\left(\frac N2+1\right).
\label{eq:egvalmaxmin}
\end{equation}
where $N_p$ is the number of neutrinos at the oscillation frequency $\omega_p$, and where we have suppressed the radial dependence in the notation of $\mu$ for convenience. The remaining eigenstates of the Hamiltonian may then be systematically constructed by first defining the Gaudin algebra with the operators 
\begin{equation}
\vec{S}(\zeta_\alpha) \equiv \sum_{p}\frac{\vec{J}_p}{\omega_p-\zeta_\alpha}
\label{gaudin}
\end{equation}
where we have introduced a sequence of Bethe Ansatz variables $\zeta_\alpha$, which are yet to be determined, following the formalism introduced in \cite{Pehlivan:2011hp}.

Our Bethe Ansatz is the claim that the eigenstates of our system are states of the form
\begin{equation}
\ket{\zeta_1,\ldots,\zeta_\kappa} = \mathcal N(\zeta_1,\ldots,\zeta_\kappa) \bigg(\prod_{\alpha=1}^\kappa S^{-}_\alpha \bigg)\ket{j,+j}
\label{eigenstates}
\end{equation}
where $\ket{j,+j}$ is an eigenstate, $S_\alpha^-=S^-(\zeta_\alpha)$, and $\mathcal N(\zeta_1,\ldots,\zeta_\kappa)$ is a normalization factor. An eigenstate of the Hamiltonian, such as the one defined in Eq.~\eqref{eigenstates}, is typically a linear superposition of several mass basis states (i.e., Kronecker-products of mass states of individual neutrinos), each consisting of $N/2 + j - \kappa$ neutrinos in the $\nu_1$ state and $N/2 - j + \kappa$ neutrinos in the $\nu_2$ state. The fact that each eigenstate has a well-defined number of $\nu_1$ and $\nu_2$ is a consequence of the total $J^z = \sum_p J_p^z$ commuting with the Hamiltonian $H$ from Eq.~\eqref{eq:saham}. Eigenstates of the Hamiltonian are therefore also eigenstates of $J^z$:
\begin{equation}
J^z \ket{\zeta_1,\ldots,\zeta_\kappa} = \left( j - \kappa \right) \ket{\zeta_1,\ldots,\zeta_\kappa}.
\label{eigenJz}
\end{equation}

A similar procedure may be carried out in which we instead apply $S^+_\alpha$ operators to $\ket{j,-j}$ to obtain such eigenstates. In fact, these two procedures yield identical results. This fact can be shown by defining a rotation operator $T=e^{i\frac{\pi}{2}(J^++J^-)}$, for which $TJ^zT^{-1}=-J^z$ and $TJ^{\pm}T^{-1}=J^{\mp}$, and so $T\ket{j,-j}=\ket{j,+j}$. Moreover, one can see that, if $\ket{\psi}$ is an eigenstate obtained from Eq.~\eqref{eigenstates}, then $T\ket{\psi}$ is an eigenstate of a Hamiltonian $THT^{-1}= -\sum_{p}\omega_p\vec{B}\cdot\vec{J}_p+\mu(r)\vec{J}\cdot\vec{J}$ that may be obtained from a \lq\lq raising formalism\rq\rq\ using $S^+_\alpha$ instead.

For a particular eigenstate consisting of $\kappa$ neutrinos in the isospin-down ($\nu_2$) configuration, we will have $\kappa$ different ansatz variables $\zeta_\alpha$ to determine, and one can show \cite{Birol:2018qhx,Balantekin:2018mpq} using the commutation relations between the Gaudin operators that this requirement is equivalent to the condition
\begin{equation} 
- \frac{1}{2\mu} - \sum_{p=1}^M\frac{j_p}{\omega_p-\zeta_\alpha} = \sum_{\substack{\beta=1 \\ \beta \neq \alpha}}^\kappa\frac{1}{\zeta_\alpha-\zeta_\beta},
\label{originalBA}
\end{equation}
for $\alpha=1,\ldots,\kappa$. Here, $M$ is the total number of energy (or equivalently, $\omega$) values in the system, and where $j_p$ is related to the eigenvalue of the Casimir operator $\vec{J}_p^2$ (each $j_p$ can take values from $0$ or $1/2$ to $N_p/2$). For a particular choice of $\kappa$, and $j_p$s, Eq.~\eqref{originalBA} can admit multiple solutions of the form $\{\zeta_1, \ldots,\zeta_\kappa\}$. Obtaining the complete set of eigenstates and eigenvalues for an $N$-neutrino system therefore involves solving multiple sets of equations of the form Eq.~\eqref{originalBA}, for $\kappa = 1,\ldots,N$ and each $j_p = 0\text{ or }1/2,\ldots,N_p/2$. For the specific case where $j_p = 1/2$ for all $p$, the number of solutions of Eq.~\eqref{originalBA} for a given $\kappa$ is equal to $^NC_\kappa$, and the total number of solutions is $2^N$, across all values of $\kappa$.

One can always choose $j = N/2$ as the starting point for the construction of eigenstates using the Bethe Ansatz method. With this choice, the energy eigenvalue for the eigenstate defined in Eq.~\eqref{eigenstates} can be shown to be 
\begin{equation}
E=E_{+N/2}+\sum_{\alpha=1}^\kappa \zeta_\alpha-\mu \kappa(N-\kappa+1).
\label{eigenvalues}
\end{equation}

\section{The Lambda method} \label{sec:Lambdaeq}

The Bethe Ansatz equations, Eq.~\eqref{originalBA}, constitute a set of coupled algebraic equations in $\kappa$ variables. In principle, numerical solutions to this set of equations may be sought\----however, in their original form the equations are unwieldy. For starters, the variables $\zeta_\alpha$ admit complex values, and the equations contain singularities for certain values of parameters $\omega_p$ and $\mu$ where the different $\zeta_\alpha$ approach each other. Moreover, if each equation were to be converted into a coupled polynomial form (by cross-multiplying all the denominators), then the order of each polynomial would be $M + \kappa - 2$. Therefore, it is worthwhile to explore the possibilities of recasting the Bethe Ansatz equations  into a different, more tractable form.

In order to accomplish this, one may introduce certain auxiliary functions which depend on the Bethe Ansatz variables. For instance, following Refs.~\cite{Faribault:2011rv,Babelon_2007} we may define the function 
\begin{equation}
\Lambda(\lambda)=\sum_{\alpha=1}^\kappa\frac{1}{\lambda-\zeta_\alpha},
\label{Lambda}
\end{equation}
and transform Eq.~\eqref{originalBA} into a first-order ordinary differential equation
\begin{equation}
\Lambda(\lambda)^2+\Lambda'(\lambda)+\frac{1}{\mu}\Lambda(\lambda)=\sum_{q=1}^M2j_q\frac{\Lambda(\lambda)-\Lambda(\omega_q)}{\lambda-\omega_q},
\label{eq:LambdaDE}
\end{equation}
where the prime represents derivative with respect to $\lambda$. Eq.~\eqref{eq:LambdaDE} is not straightforward to integrate because of the presence of the parameters $\Lambda(\omega_q)$, whose values are not known \textit{a priori}, and are in fact dependent on the equation itself. These parameters can be determined by taking the limit of Eq.~\eqref{eq:LambdaDE} as $\lambda\to\omega_p$ for each $p=1,\ldots, M$. Doing so yields the following system of equations:
\begin{equation}
\Lambda_p^2+(1-2j_p)\Lambda_p'+\frac{1}{\mu}\Lambda_p=\sum_{\substack{q=1 \\ q\neq p}}^M2j_q\frac{\Lambda_p-\Lambda_q}{\omega_p-\omega_q},
\label{LambdaPDEs}
\end{equation}
where $\Lambda_p = \Lambda(\omega_p)$ and $\Lambda_p' = \Lambda'(\omega_p)$, for $p=1,\ldots, M$. % Observe that $\Lambda_p$ is a variable that may depend on not only $\omega_p$ but also other $\omega_q$ ($q\neq p$). 
In particular, if $j_p=1/2$ for all $p$, then our equations for $\Lambda_p$ reduce simply to the form
\begin{equation}
\Lambda_p^2 + \frac{1}{\mu}\Lambda_p= \sum_{\substack{q=1 \\ q\neq p}}^M \frac{\Lambda_p-\Lambda_q}{\omega_p-\omega_q},
\label{algebraicLambda}
\end{equation}
yielding a system of coupled algebraic equations of quadratic order in the parameters $\Lambda_p$. Physically, this represents the case where the system is composed of neutrinos that all have pairwise distinct momenta, allowing us to choose a discrete set of $\omega_p$ bins in which each bin includes exactly one neutrino. Details of the derivation of Eqs.~\eqref{eq:LambdaDE} and \eqref{algebraicLambda} are given in Appendix \ref{sec:applambda}.

Eq.~\eqref{algebraicLambda} is manifestly much simpler than the original Bethe Ansatz equations Eq.~\eqref{originalBA}. Moreover, unlike Eq.~\eqref{originalBA} where each value of $\kappa = 1,\ldots,N$ requires solving a separate set of Bethe Ansatz equations, Eq.~\eqref{algebraicLambda} represents just a single set of equations that can be solved to yield all the solutions corresponding to different values of $\kappa$. In Sec.~\ref{sec:rootsfromLambda}, we show that the following relation holds between the variables $\Lambda_p$ and $\zeta_\alpha$:
\begin{equation}
\sum_{p=1}^M j_p\omega_p\Lambda_p = \kappa\sum_{p=1}^Mj_p-\frac{1}{2\mu}\sum_{\alpha=1}^\kappa \zeta_\alpha-\frac{\kappa(\kappa-1)}{2},
\label{LambdaX}
\end{equation}
which may be used to express our energy eigenvalues from Eq.~\eqref{eigenvalues} in terms of $\Lambda_p$ instead of the Bethe Ansatz variables $\zeta_\alpha$:
\begin{equation}
E = E_{N/2}-2\mu\sum_{p=1}^M j_p\omega_p\Lambda_p.
\label{eigenvaluesLambda}
\end{equation}

Thus, we seek to determine the parameters $\Lambda_1,\ldots,\Lambda_M$ from Eq.~\eqref{algebraicLambda}. Following that, one may use Eq.~\eqref{Lambda}, which may then be inverted to obtain the Bethe Ansatz variables $\zeta_\alpha$. The $\zeta_\alpha$ may then be used to reconstruct the states and their energies that solve our model using Eqs.~\eqref{gaudin} and \eqref{eigenvalues}. Alternatively, we show in this paper that for the system in which $j_p=1/2$ for all $p$, one may instead directly reconstruct the eigenstates and their energies using the variables $\Lambda_p$, without involving the Bethe Ansatz variables. These two procedures are detailed in Secs.~\ref{sec:rootsfromLambda} and \ref{sec:eiglambda}, respectively.

\subsection{Obtaining the Bethe Ansatz roots \texorpdfstring{$\zeta_\alpha$}{zeta} from the variables \texorpdfstring{$\Lambda_p$}{Lambdap}} \label{sec:rootsfromLambda}

%For convenience of notation, let us henceforth define $\Lambda_p \equiv \Lambda(\omega_p)$, $\Lambda'_p = \Lambda'(\omega_p)$, and likewise for higher order derivatives. \\
After solving Eq.~\eqref{algebraicLambda} for $\Lambda_p$, it is possible to reduce the problem of obtaining the Bethe Ansatz variables $\zeta_\alpha$ to that of solving a single polynomial equation of order $\kappa$. This process involves two steps, the first being deriving a set of constraint relations between the $\Lambda_p$s and the power sums of $\zeta_\alpha$s.
%It is possible to derive a set of constraint equations involving $\{\Lambda_p\}$ and $\{\zeta_\alpha\}$. 
Using the definition Eq.~\eqref{Lambda}, one has
\begin{equation}
\begin{split}
\sum_{p=1}^{M} j_p \, \omega_p^k \, \Lambda_p &= \sum_{\alpha=1}^{\kappa} \sum_{p=1}^M j_p \frac{\omega_p^k}{\omega_p - \zeta_\alpha} \\
	&= \sum_\alpha \sum_p j_p \frac{\omega_p^k - \zeta_\alpha^k}{\omega_p - \zeta_\alpha} + \sum_\alpha \zeta_\alpha^k \sum_p \frac{j_p}{\omega_p - \zeta_\alpha} \\
	&= \sum_\alpha \sum_p j_p \left(\sum_{l=1}^k \omega_p^{k-l}\,\zeta_\alpha^{l-1}\right) \\
	&\qquad - \sum_\alpha \zeta_\alpha^k \sum_{\beta \neq \alpha} \frac{1}{\zeta_\alpha - \zeta_\beta} - \sum_\alpha \frac{\zeta_\alpha^k}{2\mu}.
\end{split}
\end{equation}

Here, in the final step, we have used the Bethe Ansatz equation Eq.~\eqref{originalBA} to replace the second inner sum.  Using symmetry arguments and changing the order of the summations, the above equation becomes
\begin{equation}
\begin{split}
\sum_{p=1}^M j_p \, \omega_p^k \, \Lambda_p
	&= \sum_{l=1}^k \left( \sum_{\alpha=1}^\kappa \zeta_\alpha^{l-1} \right) \left( \sum_{p=1}^M j_p \,\omega_p^{k-l} \right) \\
	&\qquad - \frac{1}{2\mu} \sum_{\alpha=1}^\kappa \zeta_\alpha^k - \frac12 \sum_{\substack{\alpha,\beta=1 \\ \alpha \neq \beta}}^\kappa \frac{\zeta_\alpha^k - \zeta_\beta^k}{\zeta_\alpha - \zeta_\beta} \\
	&= \sum_{l=1}^k \left( \sum_\alpha \zeta_\alpha^{l-1} \right) \left( \sum_p j_p \,\omega_p^{k-l} \right) \\
	&\qquad - \frac{1}{2\mu} \sum_\alpha \zeta_\alpha^k - \frac12 \sum_{\substack{\alpha,\beta \\ \alpha \neq \beta}} \sum_{l=1}^k \zeta_\alpha^{k-l}\,\zeta_\beta^{l-1}.
\end{split}
\end{equation}

Here, it is useful to define the power sums of the Bethe Ansatz variables, $Q_k \equiv \sum_{\alpha=1}^\kappa \zeta_\alpha^k$. This allows us to write the above expression as
\begin{equation} \label{eq:calcQk}
\begin{split}
\sum_{p=1}^M j_p \, \omega_p^k \, \Lambda_p
% 	&= - \frac{1}{2\mu} Q_k + \sum_{l=1}^k Q_{l-1} \left( \sum_p j_p \,\omega_p^{k-l} \right)  \\
% 	&\qquad - \frac{1}{2} \sum_{l=1}^k \left[\sum_{\alpha,\beta} \zeta_\alpha^{k-l}\,\zeta_\beta^{l-1} - \sum_\alpha \zeta_\alpha^{k-1}\right] \\
	= - \frac{1}{2\mu} Q_k + \sum_{l=1}^k Q_{l-1} \left( \sum_{p=1}^M j_p \,\omega_p^{k-l} \right) \\
	\qquad - \frac{1}{2} \left[\sum_{l=1}^k Q_{k-l}\, Q_{l-1} -  k\,Q_{k-1} \right].
\end{split}
\end{equation}

For the first few values of $k$, the above equation takes the following forms:
\begin{align}
\sum_p j_p\,\Lambda_p &= -\frac{Q_0}{2\mu} = -\frac{\kappa}{2\mu}, \label{eq:Q0eq} \\
\sum_p j_p\,\omega_p\,\Lambda_p &= -\frac{Q_1}{2\mu} + \kappa \sum_p j_p - \frac{\kappa(\kappa-1)}{2}, \label{Q1eq}\\
\sum_p j_p\,\omega_p^2\,\Lambda_p &= -\frac{Q_2}{2\mu} + \kappa \sum_p j_p \,\omega_p + Q_1 \sum_p j_p  \nonumber \\
                                    &\hspace{90pt} - (\kappa-1)\,Q_1,
\end{align}
for $k=0,1,\text{ and } 2$, respectively. In particular, the equation for $k = 0$ may be treated as a constraint relation for the solutions $\Lambda_p$ obtained by solving Eq.~\eqref{algebraicLambda}, and can also be used to classify those solutions according to $\kappa$. Also note that the equation for $k=1$ is identical to Eq.~\eqref{LambdaX}, which can be used to express the energy eigenvalues in terms of the $\Lambda_p$, as shown earlier.%, i.e., the number of neutrinos in excited isospin states.

%\section{Obtaining the roots $\zeta_\alpha$ from the solutions $\Lambda(\omega_p)$}

Using Eq.~(\ref{eq:calcQk}), one can successively calculate the power sums $Q_k$ to any desired value of $k$, once all the $\Lambda_p$ are known. And the first $\kappa$ power sums, $Q_1, \ldots,Q_\kappa$, can be used to obtain all the roots $\zeta_\alpha$. This involves first calculating the elementary symmetric polynomials of the $\zeta_\alpha$s from the power sums. The elementary symmetric polynomials are $e_0 = 1$, $e_1 = \sum_\alpha \zeta_\alpha$, $e_2 = \sum_{{\alpha,\beta<\alpha}} \zeta_\alpha \zeta_\beta$, and so on. For $k \leq \kappa$, these can be calculated recursively from the power sums using Newton's identities:
\begin{equation}
k \,e_k(\zeta_1,\ldots,\zeta_\kappa) = \sum_{i=1}^k (-1)^{i-1} e_{k-i}\, Q_i.
\end{equation}
Therefore, one has
\begin{align}
e_1 &= e_0\, Q_1 = Q_1, \\
e_2 &= \frac{1}{2} \left( e_1\, Q_1 - e_0\, Q_2\right), 
\end{align}
and so on. Once the $\kappa$ elementary symmetric polynomials $e_1,\ldots,e_\kappa$ are evaluated, then the polynomial $P(\lambda) \equiv \prod_{\alpha=1}^\kappa (\lambda - \zeta_\alpha)$, whose roots are ${\zeta_\alpha: \alpha=1,\ldots,\kappa}$ may be constructed as
\begin{equation}
P(\lambda) = \sum_{k=0}^\kappa (-1)^k \, e_k \, \lambda^{\kappa-k}.
\end{equation}

Any one-dimensional polynomial root-finding algorithm can be employed to numerically obtain the roots $\zeta_\alpha$ of this polynomial. Once the $\zeta_\alpha$ are determined, then the eigenstates of the Hamiltonian may be explicitly constructed using Eq.~\eqref{eigenstates}. As an aside, it is also interesting to note that $P(\lambda)$ is related to the function $\Lambda(\lambda)$ as
$\Lambda(\lambda) = {P'(\lambda)}/{P(\lambda)}$.% = \sum_{i=1}^N \frac{1}{\lambda - \zeta_\alpha}$.

\subsection{Constructing the eigenstates directly using \texorpdfstring{$\Lambda_p$}{Lambdap}} \label{sec:eiglambda}

Alternatively, we show that it is possible to directly compute the eigenstates of the Hamiltonian in terms of the auxiliary variables, without needing to calculate the Bethe Ansatz variables first. Unlike the procedure described in Sec.~\ref{sec:rootsfromLambda}, this method does not involve any additional numerical root-finding after obtaining the $\Lambda_p$, and therefore eliminates a potential source of numerical error. We can use the fact $[J_p^-,J_q^-]=0$ to find that the right hand side of Eq.~\eqref{eigenstates} can be rewritten directly in terms of $\Lambda_p$ (without involving the $\zeta_\alpha$) for any $\kappa$. Explicit derivations of these identities for $\kappa=2,3$ are provided in Appendices~\ref{sec:appeig23} and \ref{sec:appeig23_2}, and here we present the argument for a generic $\kappa$. Let us define the $\kappa\times\kappa$ matrix $A$ with the matrix elements $A_{ij}=S_i^-\delta_{ij}$, where $\delta_{ij}$ is the Kronecker delta. Disregarding the normalization for the time being, one can rewrite Eq.~\eqref{eigenstates} as
\begin{equation} \label{eigstdet}
\begin{split}
\ket{\zeta_1,\ldots,\zeta_\kappa} &= e_\kappa(S_1^-,\ldots,S_\kappa^-)\ket{j,+j} \\
	&= \det(A)\ket{j,+j}, 
\end{split}
\end{equation}
where $e_\kappa$ is the $\kappa$-th elementary symmetric polynomial and where we have used the fact that $e_\kappa(x_1, \ldots, x_\kappa) = \prod_{j=1}^\kappa x_j$ for any variables $x_j$. Since $A$ is a square matrix over a commutative ring, we can use the Cayley-Hamilton Theorem to infer that
\begin{equation}
\ket{\zeta_1,\ldots,\zeta_\kappa} = \frac{1}{\kappa!}\sum_{\sigma\in\mathrm{S}(\kappa)}\mathrm{sgn}(\sigma)\mathrm{tr}_\sigma(A)\ket{j,+j}
\label{eigenstatesRewriting}
\end{equation}
where $S(\kappa)$ is the symmetry group of $\kappa$ letters and $\mathrm{tr}_\sigma(A) = \mathrm{tr}(A^{f_1}) \cdots \mathrm{tr}(A^{f_n})$ for a permutation $\sigma$ of cycle type $(f_1,\ldots,f_n)$ \cite{bourbaki1998algebra}. The traces may be expressed in terms of the $\Lambda_p$s via the following steps:
%where $S(\kappa)$ is the symmetry group of $\kappa$ letters. $\mathrm{tr}$ is the trace over the indices of $A$, $\mathrm{tr}_\sigma(A)=\mathrm{tr}(A^{f_1})\cdots\mathrm{tr}(A^{f_m})$ with $(f_1,\ldots,f_m)$ as the cycle type of the permutation  $\sigma$ \cite{bourbaki1998algebra}. Then, we can use the Heaviside ``cover-up rule'' to simplify these traces:
\begin{widetext}
\begin{align}
\mathrm{tr}(A^f) &= \sum_{\alpha=1}^\kappa(S_\alpha^-)^f = \sum_{i=1}^\kappa \sum_{p_1=1}^M\cdots\sum_{p_f=1}^M \frac{J_{p_1}^-\cdots J_{p_f}^-}{(\omega_{p_1}-\zeta_\alpha) \ldots (\omega_{p_f}-\zeta_\alpha)}
\nonumber \\
	&= \sum_{\alpha=1}^\kappa \sum_{p_1=1}^M\cdots\sum_{p_f=1}^M J_{p_1}^-\cdots J_{p_f}^- \sum_{m=1}^f \frac{1}{\omega_{p_m}-\zeta_\alpha} \prod_{\substack{l=1 \\ l\neq m}}^f \frac{1}{\omega_{p_l}-\omega_{p_m}}
\nonumber \\
	&= \sum_{p_1=1}^M\cdots\sum_{p_f=1}^M J_{p_1}^-\cdots J_{p_f}^- \sum_{m=1}^f \Lambda_{p_m} \prod_{\substack{l=1 \\ l\neq m}}^f \frac{1}{\omega_{p_l}-\omega_{p_m}}.
\label{coverup}
\end{align}
\end{widetext}
Here we have used the Heaviside cover-up rule between the second and third equalities, and the definition from Eq.~\eqref{Lambda} between the third and fourth equalities (after exchanging the order of summation). 
An alternative derivation for the individual overlaps of energy eigenstates with the mass basis states is given in Ref. \cite{Claeys:2017sp,Claeys:2018zwo}.

Since the form of $\mathrm{sgn}(\sigma)\mathrm{tr}_\sigma(A)$ depends on the cycle type\----which have multiplicities $c_\sigma$\----but not the particular $\sigma$, we can reduce Eq.~\eqref{eigenstatesRewriting} to the following:
%Since the form of $\mathrm{sgn}(\sigma)\mathrm{tr}_\sigma(A)$ depends on the cycle type -- which have multiplicities $c_\sigma$ --  but not the particular $\sigma$, we can reduce Eqn \eqref{eigenstatesLambda}
\begin{align}
\ket{\zeta_1,\ldots,\zeta_\kappa} &=\frac{1}{\kappa!}\sum_{\mathrm{cycle}\:\mathrm{types}}\mathrm{sgn}(\sigma)c_\sigma\mathrm{tr}_\sigma(A)\ket{j,+j}.
\label{eigenstatesLambdas}
\end{align}
%\begin{align}
%\ket{\zeta_1,\ldots,\zeta_\kappa} &=\frac{1}{\kappa!}\sum_{\mathrm{cycle}\:\mathrm{types}}\mathrm{sgn}(\sigma)c_\sigma\mathrm{tr}_\sigma(A)\ket{j,+j}.
%\label{eigenstatesLambdaSimplified}
%\end{align}

Lastly, we note that reversing the order of $l$ and $m$ in the denominator of Eq.~\eqref{coverup} for each factor $\mathrm{tr}(A^{f})$ in $\mathrm{tr}_\sigma(A)$ produces a factor of $\mathrm{sgn}(\sigma)$, which cancels with the same factor already written in each term of Eq.~\eqref{eigenstatesRewriting}. In fact, using a decomposition into cycle-types, $\sigma=(f_1,\ldots,f_n)$, we see that $\mathrm{sgn}(\sigma)=\prod_{i=1}^n(-1)^{f_i+1}=(-1)^{\kappa+n}$, while a factor of $\prod_{i=1}^n(-1)^{f_i-1}=(-1)^{\kappa-n}$ is produced from reversing the differences mentioned in the previous sentence.\footnote{Interestingly, the resulting form of Eq.~\eqref{eigenstatesRewriting} with this cancellation suggests that the module spanned by $S_1^-,\ldots,S_\kappa^-$ forms a $\kappa$-th-power symmetric (as opposed to exterior) algebra, whose character can be described with the same formula.}

Alternatively, since $\mathrm{tr}(A^f)$ is simply the power sum $\sum_{i=1}^\kappa(S_i^-)^f $, one may use Newton's identities to systematically construct the elementary symmetric polynomial $e_\kappa(S_1^-,\ldots,S_\kappa^-)$ using the traces $\mathrm{tr}(A),\ldots,\mathrm{tr}(A^\kappa)$. This was the basis of our numerical approach to calculating the eigenstates, the results of which are shown in Sec.~\ref{sec:bethesol}.

\section{Solutions of the Bethe Ansatz equations} \label{sec:bethesol}

Having obtained a set of algebraic equations in the auxiliary variables $\Lambda_p$, we are now in a position to discuss the solutions to these equations. To begin with, one can examine the solutions in the limit $\mu \to 0$, which we henceforth shall also refer to as the \lq\lq asymptotic limit.\rq\rq\ At this point, it is instructive to define the quantities $\widetilde{\Lambda}_p \equiv \mu\Lambda_p$ to rewrite Eq.~\eqref{algebraicLambda} as 
\begin{equation}
\widetilde{\Lambda}_p^2+\widetilde{\Lambda}_p=\mu\sum_{\substack{q=1\\q\neq p}}^M\frac{\widetilde{\Lambda}_p-\widetilde{\Lambda}_q}{\eta_{pq}}.
\label{tildeLambda}
\end{equation}

For convenience, we have defined $\eta_{pq}=\omega_p-\omega_q$. Taking the limit $\mu \to 0$ decouples the various modes from each other, yielding the solutions $\widetilde\Lambda_p = 0$ or $-1$, independently for each $p$. These asymptotic solutions can serve as a starting point for efficiently calculating numerical solutions at a generic $\mu > 0$. This also makes it easy to see that for the case where $j_p = 1/2$ for all $p$, $2^M$ solutions exist in total. In fact, one can infer that the number of solutions is $2^M$ even when $\mu > 0$, by noting that Eq.~\eqref{tildeLambda} is a set of $M$ mutually co-prime coupled quadratic equations in $M$ variables, and therefore has a \emph{finite} solution set. One can then invoke Homotopy continuation to argue that each solution for a generic $\mu > 0$ is continuously connected to a unique solution in the $\mu \to 0$ limit~\cite{Garcia1979} (see also, Appendix~\ref{sec:apphomotopy}).

\subsection{Algebraic Solutions for \texorpdfstring{$M=2$}{M=2}} \label{sec:algebraicM2}

For the purposes of gaining some mathematical and physical intuition, we first show the analytic solutions for a simple system consisting only of two interacting neutrinos, at frequencies $\omega_1$ and $\omega_2$. This corresponds to solving Eq.~\eqref{algebraicLambda} for $M = 2$ and can serve as a test bed for the numerical technique that we later implement for dealing with the $M > 2$ cases, for which algebraic solutions either do not exist or are difficult to obtain. 

\begin{table*}[htb]
\centering
\caption{Algebraic solutions for $\Lambda_p$ from Eq.~\eqref{algebraicLambda} with $M=2$.}
\vspace{1mm}
\begin{tabular}{| l | c | c |}	\hline
	&	$\Lambda_1$	&	$\Lambda_2$	\\ \hline
	$\kappa=0$	&	0	&	0	\\ \hline
	\multirow{2}{*}{$\kappa=1$} & $\displaystyle-\frac{1}{2\mu}+\frac{1}{\eta_{12}}-\text{sgn}(\eta_{12})\sqrt{\frac{1}{\eta_{12}^2}+\frac{1}{4\mu^2}}$	&	$\displaystyle-\frac{1}{2\mu}-\frac{1}{\eta_{12}}+\text{sgn}(\eta_{12})\sqrt{\frac{1}{\eta_{12}^2}+\frac{1}{4\mu^2}}$	\\	\cline{2-3}
	&	$\displaystyle-\frac{1}{2\mu}+\frac{1}{\eta_{12}}+\text{sgn}(\eta_{12})\sqrt{\frac{1}{\eta_{12}^2}+\frac{1}{4\mu^2}}$	&	$\displaystyle-\frac{1}{2\mu}-\frac{1}{\eta_{12}}-\text{sgn}(\eta_{12})\sqrt{\frac{1}{\eta_{12}^2}+\frac{1}{4\mu^2}}$	\\	\hline
	$\kappa=2$	&	$\displaystyle-\frac{1}{\mu}$	&	$\displaystyle-\frac{1}{\mu}$	\\	\hline
\end{tabular}
\label{n=2algebraic}	
\end{table*}

\begin{table*}[htb]
\centering
\caption{Energy eigenvalues and eigenstates determined from Eq.~\eqref{algebraicLambda} $M=2$.}
\vspace{1mm}
\begin{tabular}{| l | c | c |}	\hline
	&	$E$	&	$\ket{E}$	\\ \hline
	$\kappa=0$	&	$\displaystyle 2\mu-\frac{1}{2}(\omega_1+\omega_2)$	&	$\ket{\nu_1,\nu_1}$	\\ \hline
	\multirow{2}{*}{$\kappa=1$} & $\displaystyle -(\omega_1+\omega_2)-\mu\bigg[1+\sqrt{1+\frac{(\omega_1-\omega_2)^2}{4\mu^2}}\bigg]$	&	$\displaystyle \frac{1}{\mathcal N_{1,+}}\big(\Lambda_{1,+}\ket{\nu_2,\nu_1}+\Lambda_{2,+}\ket{\nu_1,\nu_2}\big)$	\\	\cline{2-3}
	&	$\displaystyle -(\omega_1+\omega_2)-\mu\bigg[1-\sqrt{1+\frac{(\omega_1-\omega_2)^2}{4\mu^2}}\bigg]$	&	$\displaystyle \frac{1}{\mathcal N_{1,-}}\big(\Lambda_{1,-}\ket{\nu_2,\nu_1}+\Lambda_{2,-}\ket{\nu_1,\nu_2}\big)$	\\	\hline
	$\kappa=2$	&	$\displaystyle 2\mu+\frac{1}{2}(\omega_1+\omega_2)$	&	$\ket{\nu_2,\nu_2}$	\\	\hline
\end{tabular}
\label{n=2solns}	
\end{table*}

We thus aim to determine each $\Lambda_p$ for the model with $M=N=2$ with $j_1=j_2=1/2$. These $\Lambda_p$ can easily be found analytically from Eq.~\eqref{algebraicLambda}. The four solutions are listed in Table \ref{n=2algebraic}, and categorized by their corresponding $\kappa$ values. For a particular solution $\{\Lambda_p: p=1,\ldots,M\} $, $\kappa$ can be determined using the identity
\begin{equation}
\sum_{p=1}^M \Lambda_p=-\frac{\kappa}{\mu},
\label{kappa}
\end{equation}
which was derived in Sec.~\ref{sec:rootsfromLambda} (Eq.~\eqref{eq:Q0eq}). Note that, across a complete set of solutions, $\kappa$ takes all values from $0,1,\ldots,N$, and Eq.~\eqref{kappa} holds for arbitrary $M=N$.

Furthermore, using the algebraic $\Lambda_p$ solutions from Table \ref{n=2algebraic}, we may compute the energy eigenvalues using Eqs.~\eqref{eq:egvalmaxmin} and \eqref{eigenvaluesLambda} and eigenstates using Eqs.~\eqref{eigenstatesRewriting} and \eqref{coverup} (or more specifically, Eq.~\eqref{kappa2eigenstates} for $M=2$). These eigenvalues and eigenstates are aggregated in Table \ref{n=2solns}. Note that the $\Lambda_{1,\pm}$ and $\Lambda_{2,\pm}$ referenced in the latter table for $\kappa=1$ are respectively from the two solutions in Table \ref{n=2algebraic} for $\kappa=1$. Additionally, normalization coefficients calculated for states with $\kappa=1,2$ are given by
\begin{gather}
	|\mathcal N_{1,\pm}|^2 = 4\bigg[\bigg(\frac{1}{4\mu^2}+\frac{1}{\eta_{12}^2}\bigg)\mp\frac{1}{|\eta_{12}|}\sqrt{\frac{1}{4\mu^2}+\frac{1}{\eta_{12}^2}}\bigg] \label{n=2norm1} \\
	|\mathcal N_2|^2 = \frac{1}{\mu^4} \label{n=2norm2} 
\end{gather}
respectively. Note that even though these normalization coefficients are singular as $\mu \rightarrow 0$, the eigenstates themselves are not. 

\subsection{Numerical Solutions for \texorpdfstring{$M > 2$}{M > 2}}

\begin{figure*}[htbp]
\centering
\includegraphics[width=0.75\textwidth, page=1]{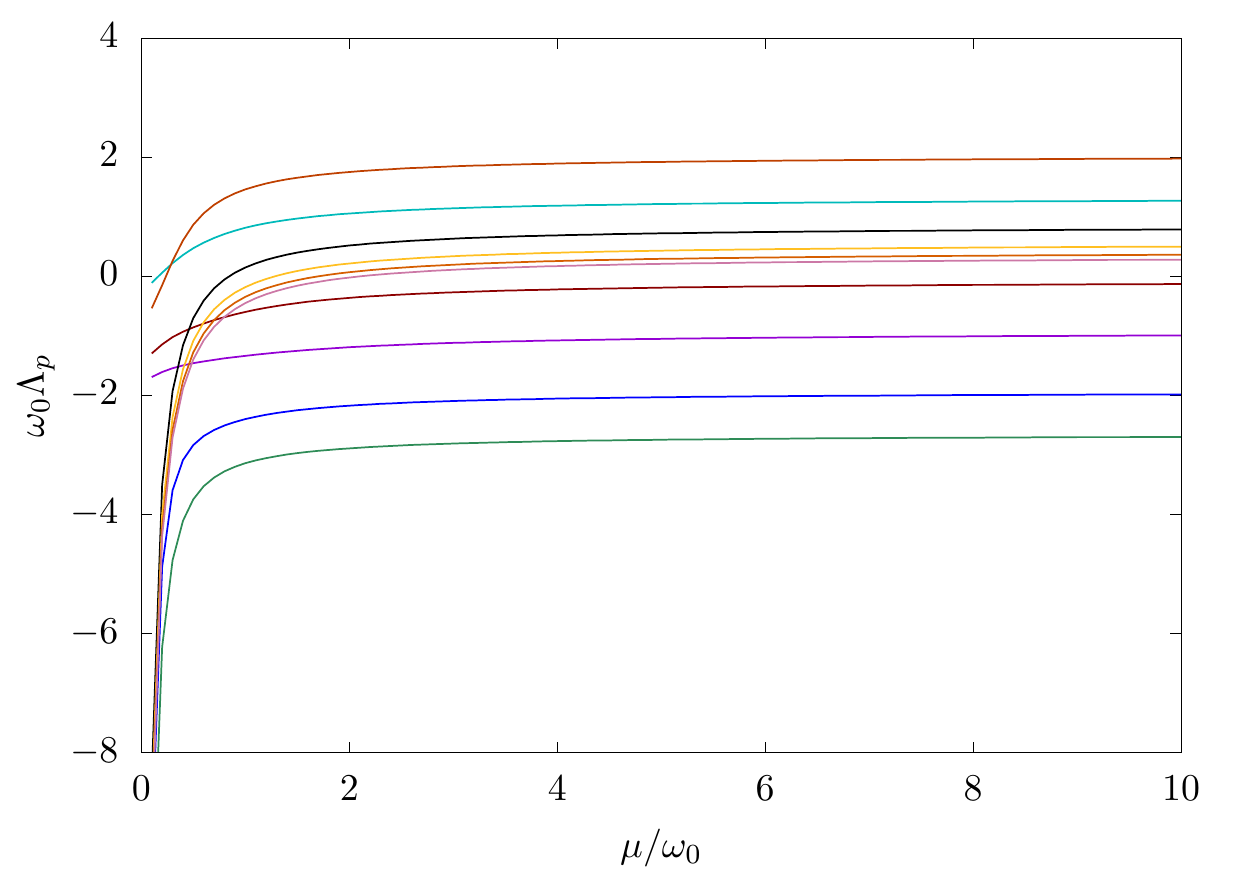} 
\caption{One out of the 210 solutions with $\kappa=6$ to Eq.~\eqref{algebraicLambda} for a system with $M=N=10$ and $\omega_p=p\omega_0$, calculated numerically using the modified Newton-Raphson method with homotopy continuation. Observe that, as $\mu \to 0$, six out of the ten $\Lambda_p$ values approach $-\infty$ (as $-1/\mu$), as expected for a $\kappa = 6$ solution.}
\label{10lambdas}
\end{figure*}

%\subsection{Homotopy continuation method for obtaining $M\geq2$ solutions}
Extending from our argument of the previous section that demonstrated there must be finitely many solutions to Eq.~\eqref{algebraicLambda} (or, equivalently, to Eq.~\eqref{tildeLambda}), we can further observe that the problem with $M > 2$ essentially necessitates solving a polynomial of degree higher than $4$ in one variable. From this observation we expect that we may not be able to algebraically solve Eq.~\eqref{algebraicLambda} or \eqref{tildeLambda}. Nevertheless, we may solve these equations numerically, as we will outline in this section.

Previously, we observed that we may obtain simple limits to the solutions of Eq.~\eqref{tildeLambda} for any $M\geq2$ as $\mu\to0$, which we refer to as the \lq\lq asymptotic solutions\rq\rq\ to our system. Moreover, we can show that the algebraic system described by Eq.~\eqref{tildeLambda} is, in a sense, homotopic (continuously deformable) to the \lq\lq asymptotic system\rq\rq\ in which we take the limit $\mu \to 0$ prior to determining its solutions. By applying a homotopy method for numerically solving algebraic systems of equations as studied in Ref. \cite{Garcia1979}, we may determine the solutions to Eq.~\eqref{tildeLambda} by tracking our solutions as they vary from the $2^M$ solutions in the limit $\mu\to0$ of our system to our desired final $\mu>0$. In fact, we build upon this existing homotopy method for our purposes: if $0\leq t\leq1$ is our homotopy parameter, then we observe that not only our solutions obtained for $t=1$, but also all solutions determined for intermediate values $0<t<1$ are physical solutions to Eq.~\eqref{tildeLambda}. A proof of the applicability of the homotopy method to our system is given in Appendix \ref{sec:apphomotopy}.

Numerical solutions can then be obtained using a generalization of the Newton-Raphson method by the homotopy method. 
We can modify this method to accommodate our homotopy process in estimating solutions to Eq.~\eqref{tildeLambda} for generic $\mu$. We apply Newton's method to solve our system of equations at the $n^\text{th}$ homotopy step:
\begin{equation}
\vec{\mathcal F}(\widetilde{\Lambda}_1^{(n)},\ldots,\widetilde{\Lambda}_M^{(n)},t^{(n)})=\vec{0},
\label{newtonhomo1}
\end{equation}
for each $t^{(n)}>0$, where $\vec{\mathcal{F}}$ is defined in Eq.~\eqref{functionH}. At the $n$-th Homotopy step, the initial guess for Newton's method is taken to be the numerical solution $\widetilde{\Lambda}_p^{(n-1)}$ obtained by solving the system of equations at the $(n-1)^\text{th}$ step, i.e., 
\begin{equation}
\vec{\mathcal F}(\widetilde{\Lambda}_1^{(n-1)},\ldots,\widetilde{\Lambda}_M^{(n-1)},t^{(n-1)})=\vec{0}.
\label{newtonhomo2}
\end{equation}

Given the form of $\vec{\mathcal F}$ in Eq.~\eqref{functionH}, we can easily compute the analytic form of the Jacobian $\partial_{\vec{z}}\vec{\mathcal F}(\vec{z}_0,t)$, to used in each step of Newton's method. Finally, all $2^M$ solutions to our system can be obtained by repeating the procedure, starting from each $\widetilde{\Lambda}_p$ independently taking values $0$ or $-1$ at $\mu=0$.

To test this method, we computed the numerical solutions to Eq.~\eqref{algebraicLambda} for $M=2$, and compared them with the analytic results described in Sec.~\ref{sec:algebraicM2}. The root-mean-squared relative error between the analytic and numerical values in comparing each component of each solution was $\lesssim10^{-14}$, suggesting that the method was able to attain a high level of numerical accuracy.

A potential issue with the utilization of homotopy continuation with large-$M$-dimensional algebraic systems is that the close proximity of solutions in $\mathbb{R}^M$ may cause numerical solvers to jump between distinct solutions as $t$ varies between 0 and 1. For larger $M$ values (such as $M=10$), introducing error-correcting algorithms during each Newton-Raphson step helps in obtaining the expected continuity of solutions as $\mu$ is varied. 

We applied this method to systems of interacting neutrinos with equally spaced oscillation frequencies given by when $\omega_p=p\omega_0$, just that only a single neutrino resides at each oscillation frequency (therefore, $j_p = 1/2$ for all $p$ and so $M=N$). Such a system of $N$ neutrinos admits $2^N$ solutions. Solutions of the system for up to $N\sim 15$ can be computed within a reasonable time frame using a personal computer. In Fig.~\ref{10lambdas}, we show the evolution with $\mu$ of one of the solutions to such a system with $N = 10$ neutrinos. Shown in the figure are the quantities $\Lambda_p = \widetilde{\Lambda}_p/\mu$, for $p = 1,\ldots,10$. 

\begin{figure*}[htbp]
\begin{center}
	\subfloat[$\kappa=0$\label{10values0}]{
		\includegraphics[width=0.48\textwidth]{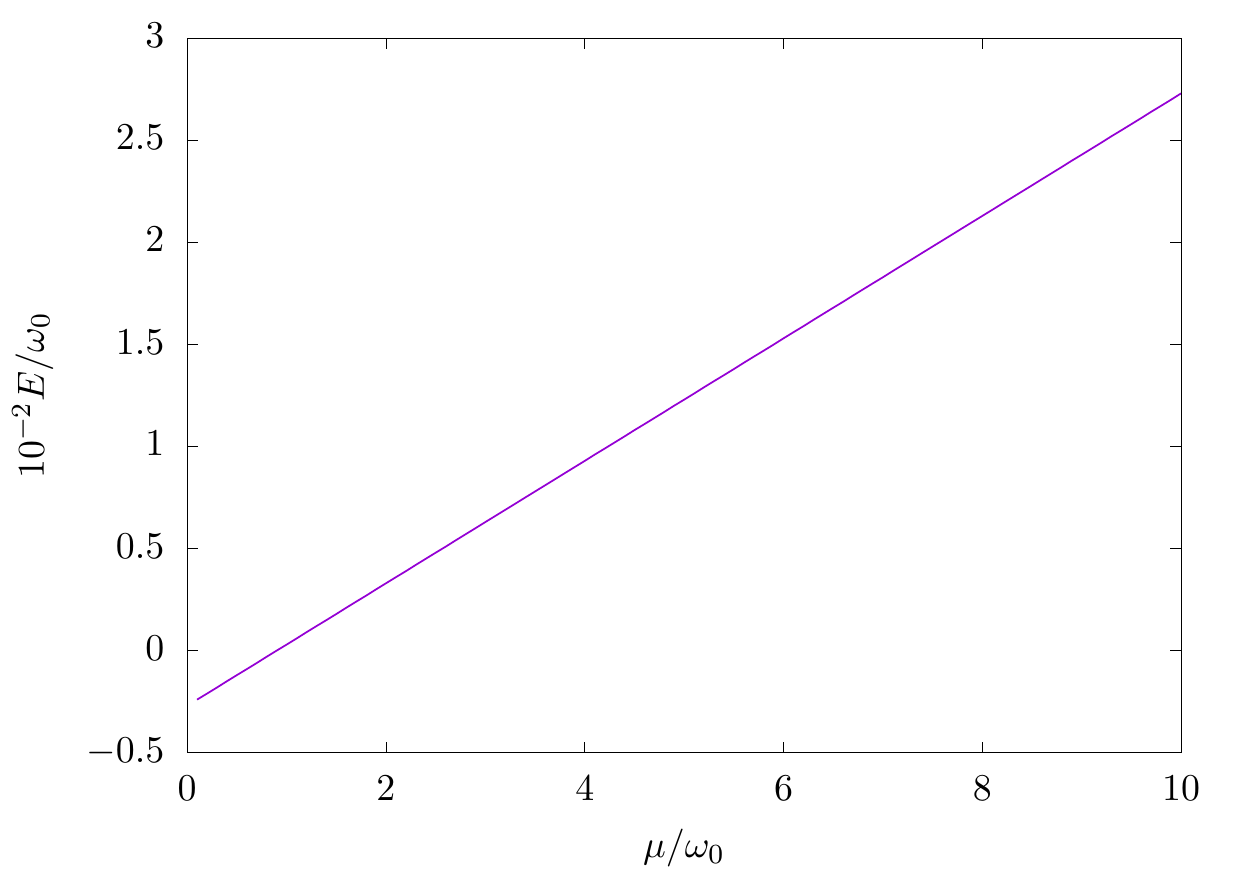}}
	~
	\subfloat[$\kappa=1$\label{10values1}]{
		\includegraphics[width=0.48\textwidth]{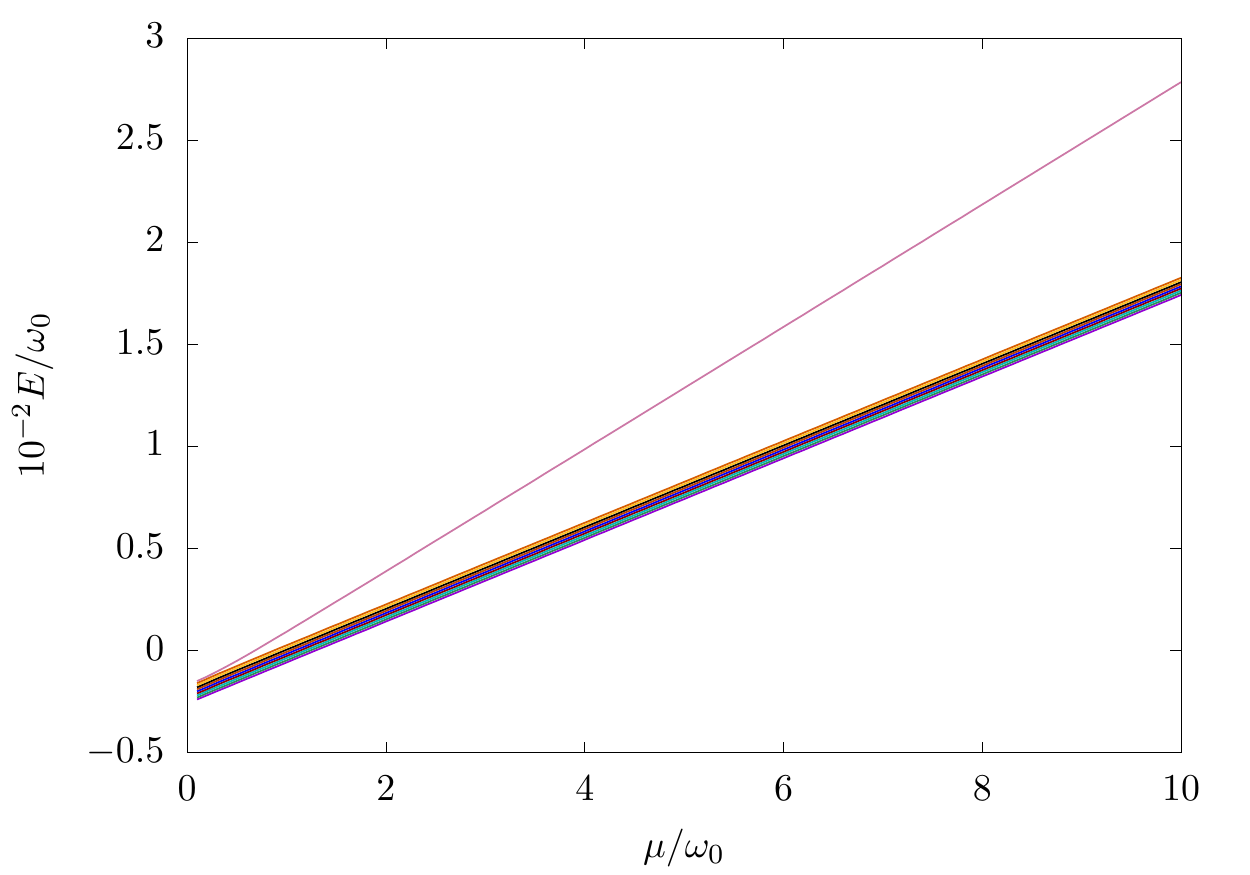}}
	
	\subfloat[$\kappa=2$\label{10values2}]{
		\includegraphics[width=0.48\textwidth]{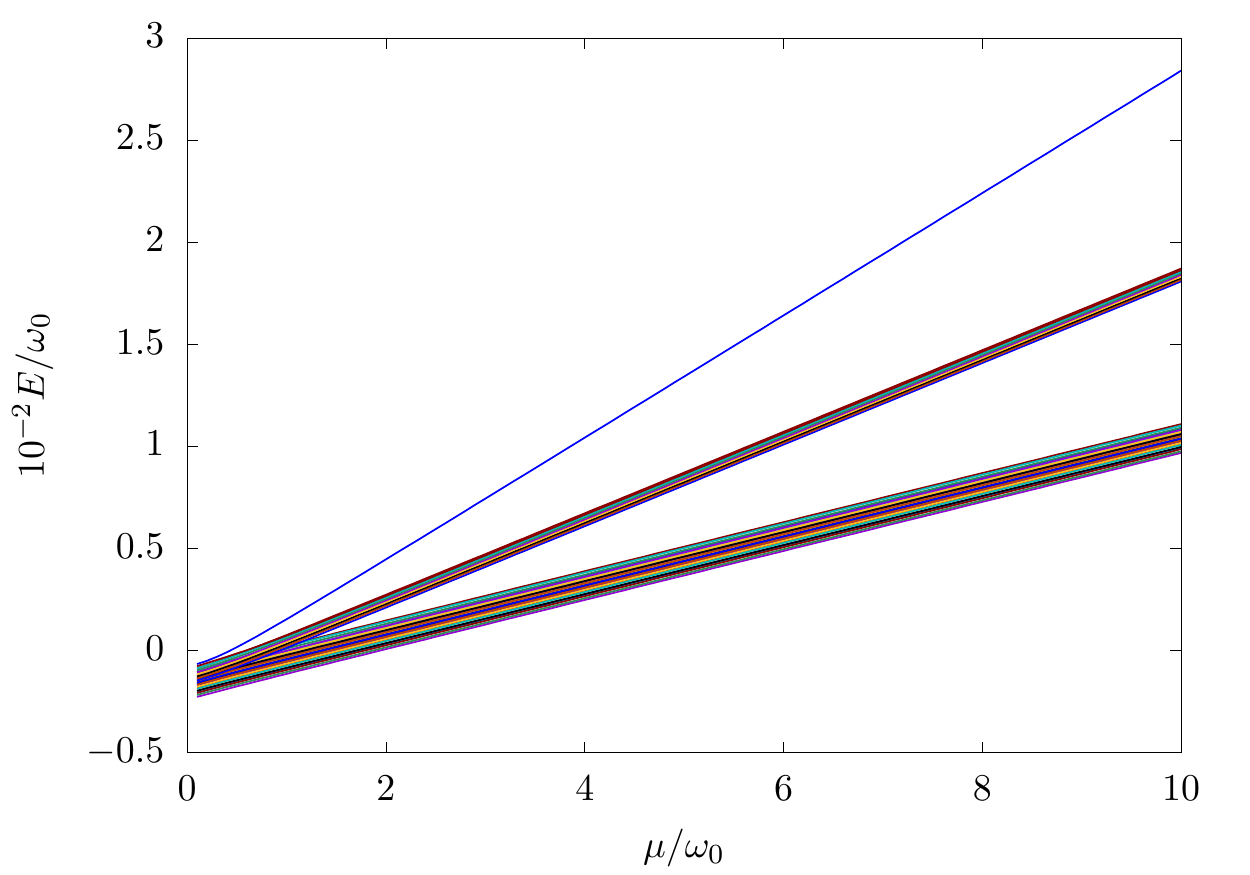}}
	~
	\subfloat[$\kappa=3$\label{10values3}]{
		\includegraphics[width=0.48\textwidth]{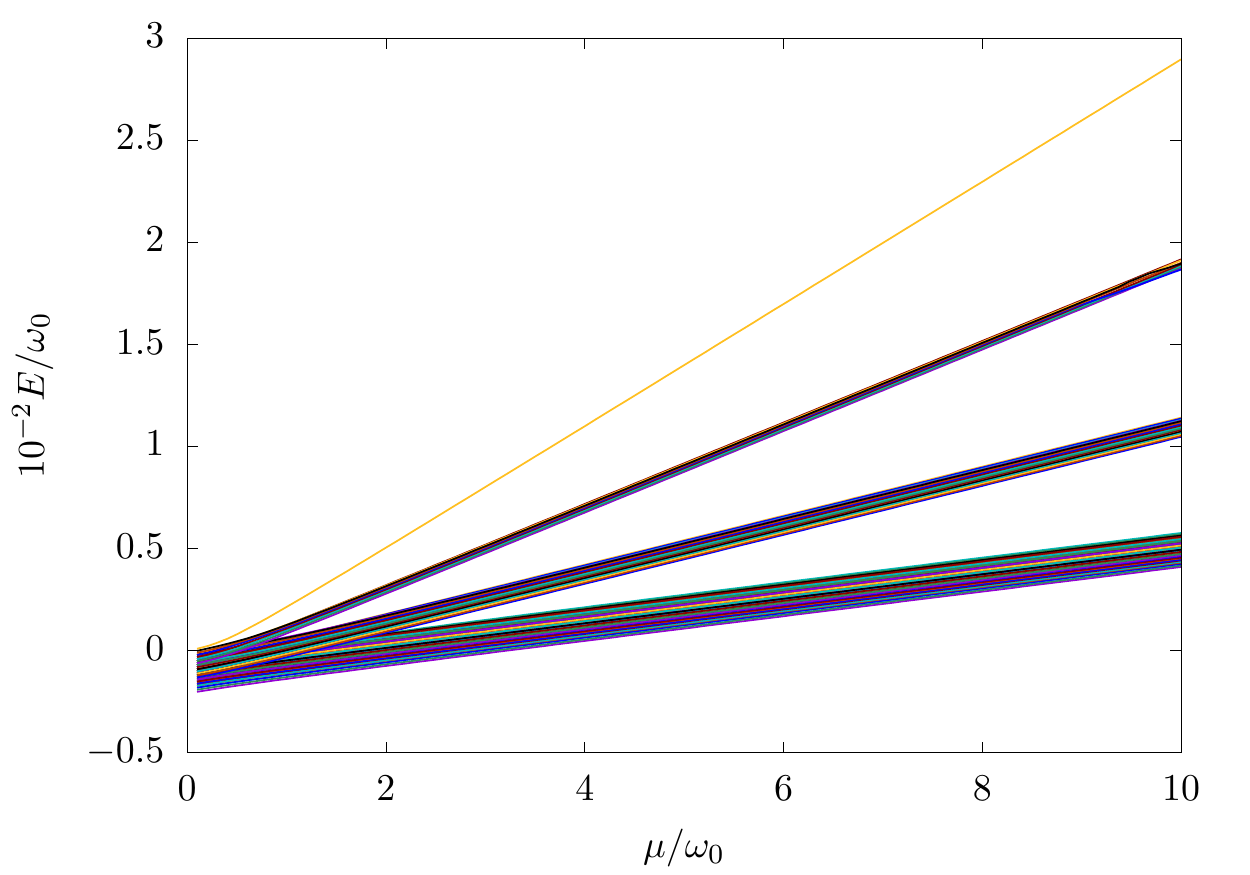}}

	\subfloat[$\kappa=4$\label{10values4}]{
		\includegraphics[width=0.48\textwidth]{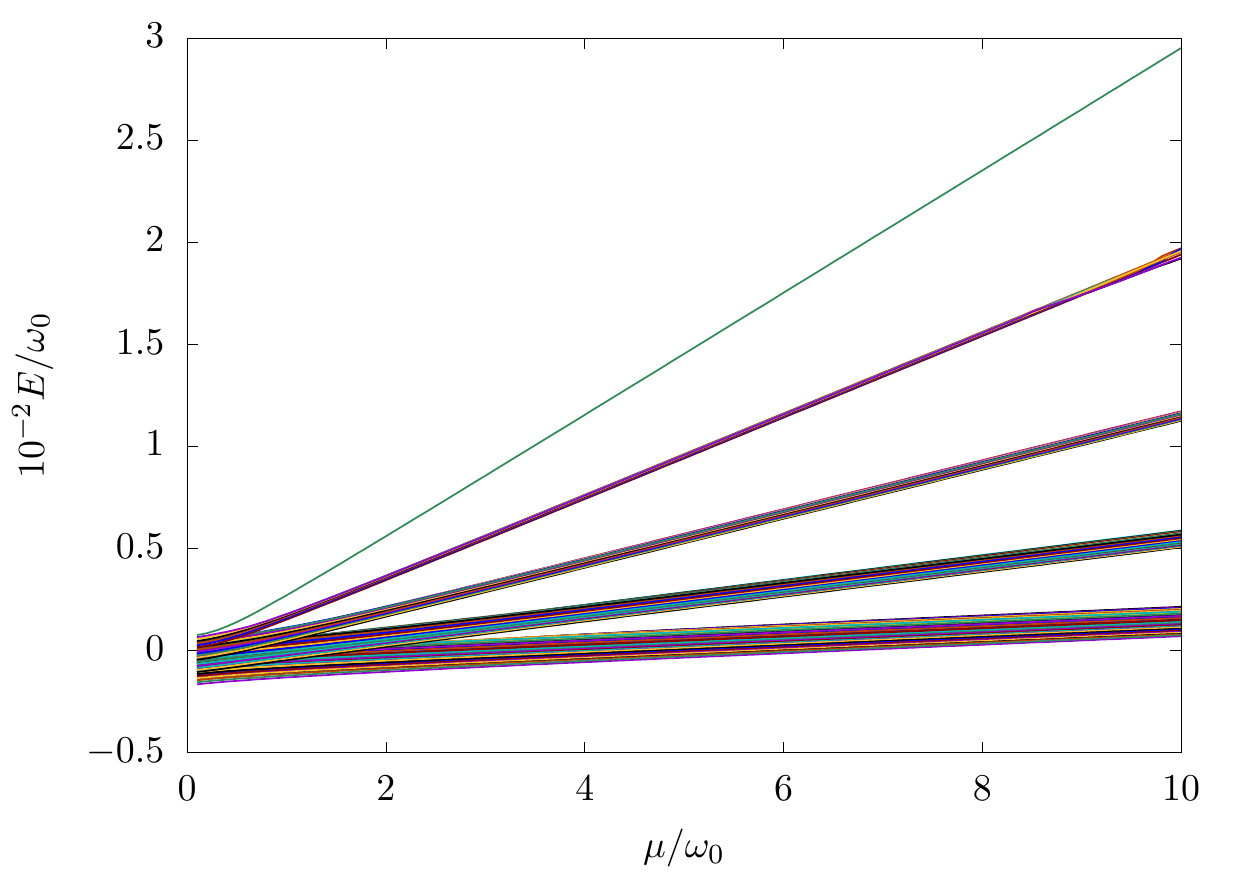}}
	~
	\subfloat[$\kappa=5$\label{10values5}]{
		\includegraphics[width=0.48\textwidth]{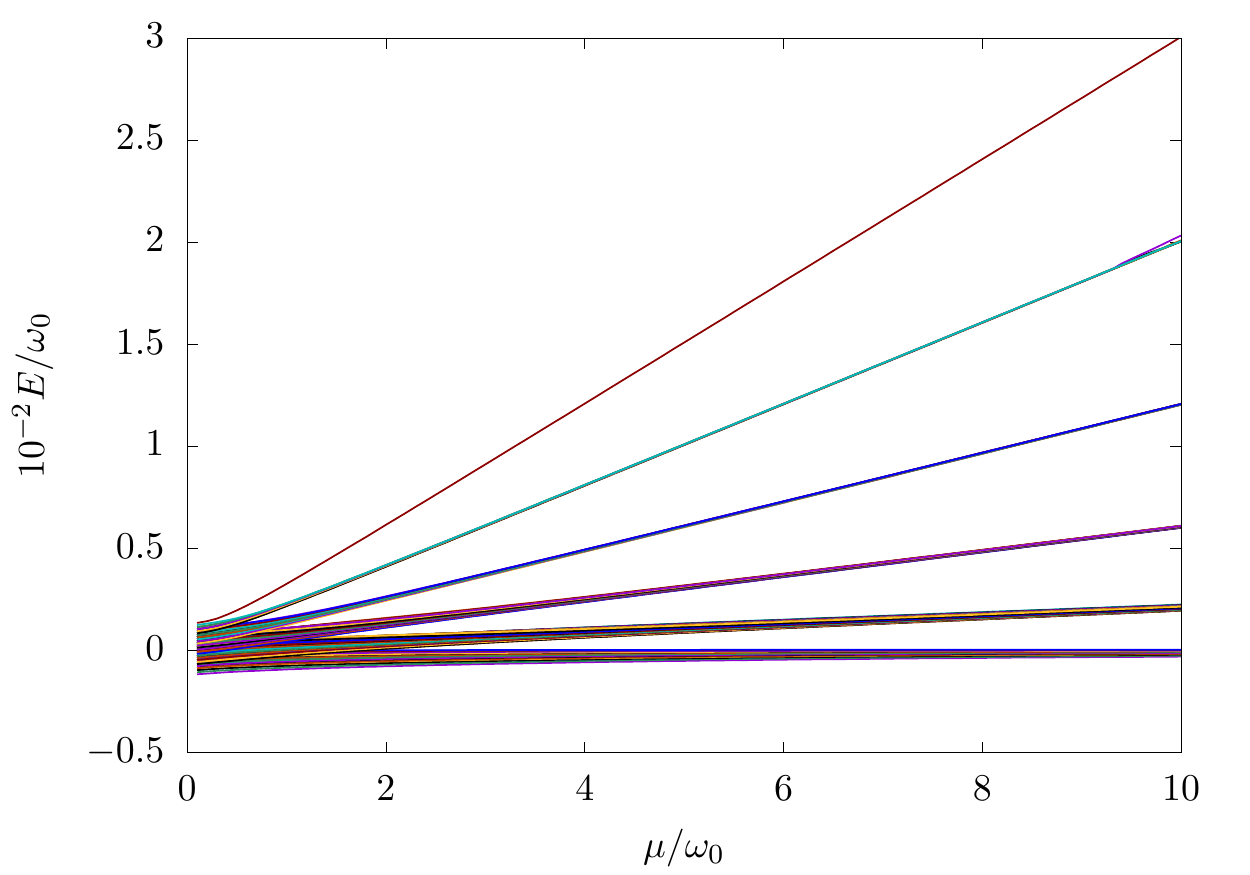}}
\end{center}
	\caption{Energy eigenvalues calculated from Eq.~\eqref{eigenvaluesLambda} after obtaining the values of each $\Lambda_p$. System parameters are identical to those in Figure \ref{10lambdas}. The solutions for $\kappa>5$ are omitted, as each of these plots are identical to their $M-\kappa$ counterparts, up to a constant vertical offset.}
	\label{10values}
\end{figure*}

We also used these solutions to calculate the eigenvalues of our Hamiltonian using Eq.~\eqref{eq:saham}. In Fig.~\ref{10values} we show all the energy eigenvalues corresponding to solutions with $\kappa = 0, \ldots, 5$, for the same system as in Fig.~\ref{10lambdas}. We also note that results for $\kappa>5$ may be obtained more efficiently by implementing a \lq\lq raising formalism\rq\rq\ analogous to our procedure from Eqs.~\eqref{eigenstates}--\eqref{eigenvaluesLambda}, in which we apply $S^+_\alpha$ operators from our Gaudin algebra to the state $\ket{j,-j}$. In analogy to $\widetilde{\Lambda}_p$, one can define the variables $\widetilde{\Lambda}_{p}^{(\uparrow)}$ in the raising formalism, which obey the coupled quadratic equations
\begin{equation}
\widetilde{\Lambda}^{(\uparrow)}_p{}^2-\widetilde{\Lambda}_p^{(\uparrow)}=\mu\sum_{\substack{q=1\\q\neq p}}^M\frac{\widetilde{\Lambda}_p^{(\uparrow)}-\widetilde{\Lambda}_q^{(\uparrow)}}{\eta_{pq}}.
\label{tildeLambdaRaise}
\end{equation}
Similar to Eq.~\eqref{kappa}, the solutions $\widetilde{\Lambda}_p^{(\uparrow)}$ obey the constraint relations
\begin{equation}
\sum_{p=1}^M j_p {\Lambda_p^{(\uparrow)}}= +\frac{\kappa^{(\uparrow)}}{2\mu}
\label{kappaRaise}
\end{equation}
where ${\Lambda_p^{(\uparrow)}} = \widetilde{\Lambda}_p^{(\uparrow)}/\mu$. In particular, a solution $\widetilde{\Lambda}^{(\uparrow)}_p$ of Eq.~\eqref{tildeLambdaRaise} can be shown to correspond to a solution $\widetilde{\Lambda}^{(\downarrow)}_p$ of Eq.~\eqref{tildeLambda}, via the identity 
\begin{equation}
    \widetilde{\Lambda}^{(\uparrow)}_p= -1 + \widetilde{\Lambda}^{(\downarrow)}_p.
    \label{duality}
\end{equation}
Using Eq. \eqref{kappa}, it can be shown that the corresponding $\kappa^{(\uparrow)}$ of a solution in the raising formalism can be related to a given $\kappa^{(\downarrow)}$ via $\kappa^{(\uparrow)}=M-\kappa^{(\downarrow)}$, further reinforcing this duality. This correspondence of solutions for a given $\kappa$ and $M-\kappa$ is also discussed in Refs.~\cite{Claeys:2017sp, Birol:2018qhx,Pehlivan:2011hp}. In particular, Ref.~\cite{Claeys:2017sp} notes that, in the limit $\mu\to\infty$, $\Lambda_p^{(\uparrow)}\to\Lambda_p^{(\downarrow)}$.

For each value of $\kappa$, one can observe that the energy eigenvalues form $b+1$ distinct branches, where $b\equiv\min\{\kappa,M-\kappa\}$, in the limit of large $\mu$, as previously noticed in \cite{Birol:2018qhx}. One may also observe that the self-interaction term in the Hamiltonian from Eq.~\eqref{eq:saham} becomes dominant as $\mu \to \infty$, and therefore the eigenstates of $H$ approximately align with the eigenstates $\ket{j,m}$ of $\vec{J}\cdot\vec{J}$, with eigenvalues $E \approx \mu j(j+1)$. For an energy eigenstate $\ket{\psi}$ with a given $\kappa$, we can see that $J^z\ket{\psi}=m\ket{\psi}$ where $m=M/2-\kappa$; therefore as $\mu \to \infty$, the quantum number $j$ of all the eigenstates for that $\kappa$ may take values in the range $M/2,\ldots,M/2-b$. So, as $\mu\to\infty$, we expect that the energy eigenvalues of states with a given $\kappa$ will split into $b+1$ branches with the approximate energies given above.

\begin{figure*}[htb]
\begin{center}
\subfloat[]{\includegraphics[width=0.49\textwidth]{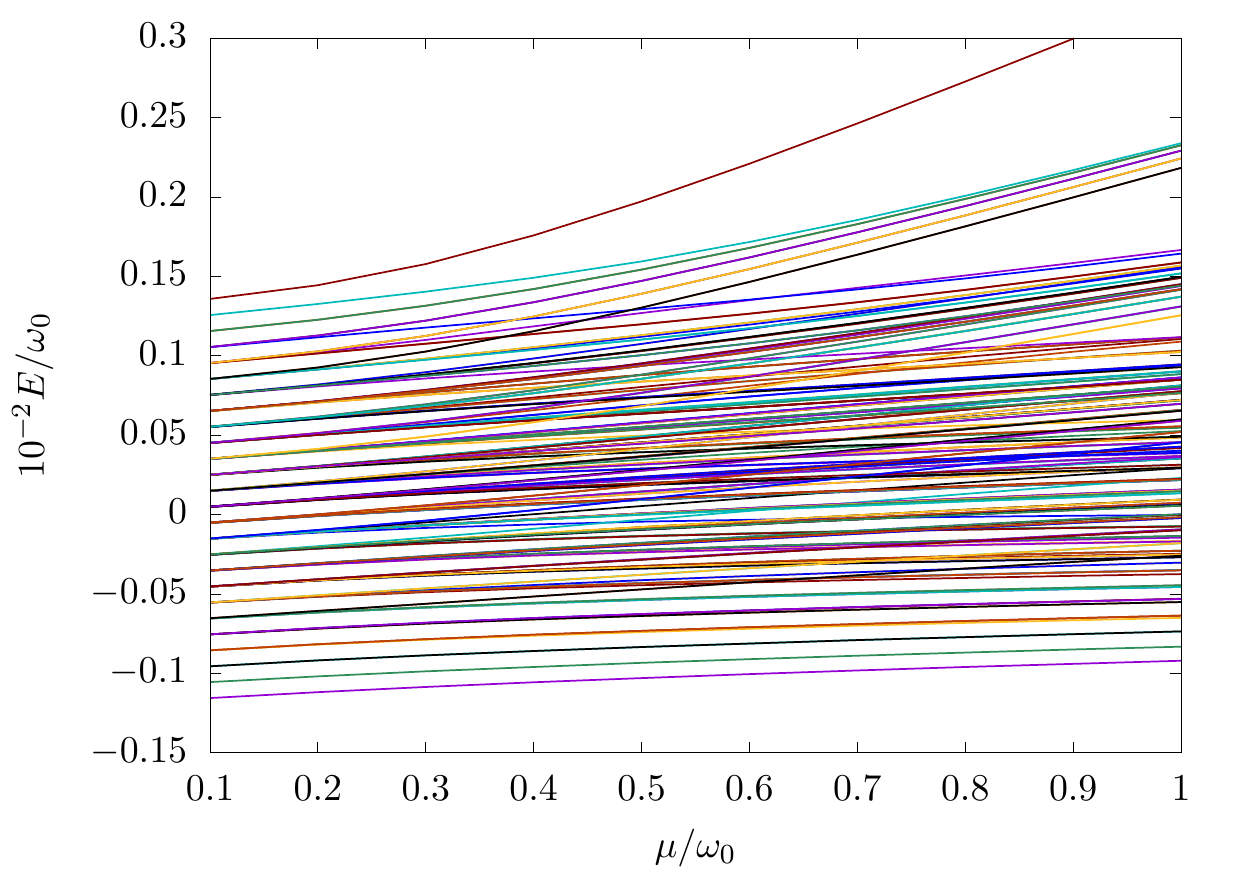} \label{10valueszoom}}
~
\subfloat[] {\includegraphics[width=0.49\textwidth]{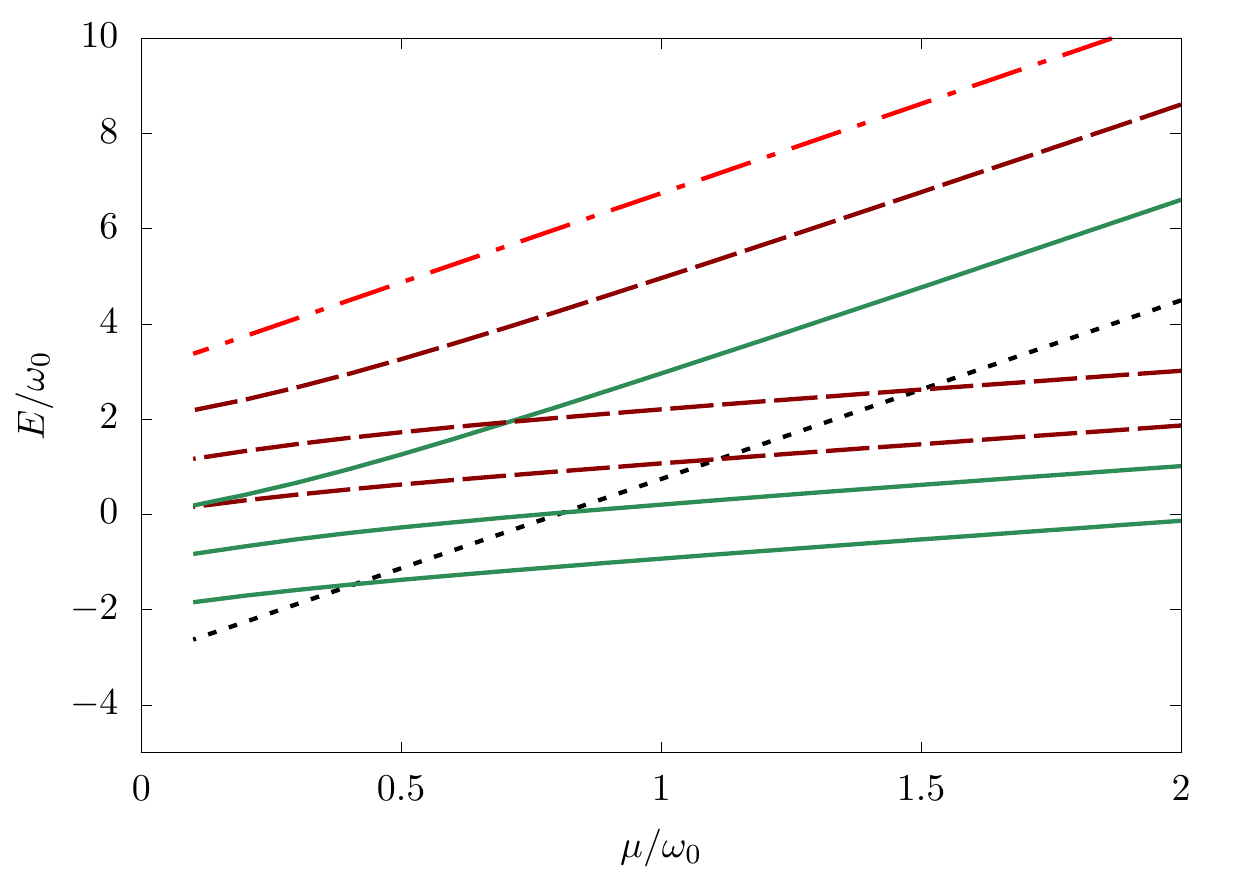} \label{3valueszoom}}
\end{center}
\caption{In the region of small $\mu$, we observe numerous level crossings between different energy states of the Hamiltonian in Eq.~\eqref{eq:saham}. Left: energy eigenvalues of all the $\kappa = 5$ solutions for the case $N = M = 10$, with the same conditions as in Figure \ref{10values5}. Right: energy eigenvalues for all the solutions of a system with $N=M=3$, with eigenvalues corresponding to different $\kappa$ coded as follows: $\kappa=0$ (dotted line), $1$ (solid lines), $2$ (dashed lines), and $3$ (dot-dashed line).}
\label{EvaluesZoom}
\end{figure*}

In Fig. \ref{10valueszoom}, we zoom in on the small-$\mu$ regions of Fig.~\ref{10values5} (all eigenvalues of the $\kappa = 5$ eigenstates) to better illustrate the various energy-level crossings that are present in this region. To examine the nature of these level crossings in better detail, we show in Fig.~\ref{3valueszoom} the energy eigenvalues of \textit{all} the states (spanning all permitted $\kappa$ values) in a system with $N = M = 3$. Eigenstates corresponding to each $\kappa$ are assigned a particular color, as described in the figure caption. Ref.~\cite{Birol:2018qhx} observes that the highest-energy states from among the solutions of each $\kappa$ (highlighted here as dashed lines) do not have energy level crossings with other states of the same $\kappa$, which is consistent with our observation. Here, we would additionally like to point out that the highest energy eigenstate of a particular $\kappa$ can cross with other eigenstates of a \textit{different} $\kappa$. However, since $J^z$ is a symmetry of the Hamiltonian, and each $\kappa$ corresponds to a unique $J^z$ eigenvalue (see Eq.~\eqref{eigenJz}), these crossings do not result in mixing between these eigenstates.

%It was assumed in Ref.~\cite{Birol:2018qhx} that the highest-energy states from among the solutions of each $\kappa$ do not exhibit energy level crossings with any other eigenstates, an assumption that was utilized in justifying the adiabatic treatment of neutrino flavor evolution from $\mu \to \infty$ all the way down to $\mu = 0$. To test this assumption, we show in Fig.~\ref{3valueszoom} the energy eigenvalues of \textit{all} the states (spanning all permitted $\kappa$ values) in a system with $N = M = 3$, and highlighted the highest energy states of $\kappa = 0,1,2,$ and $3$ with dashed lines. It is clearly observed that these states do exhibit energy level crossings\----although the highest energy state of a particular $\kappa$ does not cross with the other eigenstates of the same $\kappa$. Therefore, a more careful treatment than the one presented in Ref.~\cite{Birol:2018qhx} may be warranted.

\begin{figure*}[htb]
\begin{center}
    \subfloat[The mass-basis decomposition for the fourth excited state $\ket{\psi_2}$ in the $N=M=3$ system, as a function of $\mu$. Here, $\kappa=1$.\label{N3State}]{
	    \includegraphics[width=0.48\textwidth]{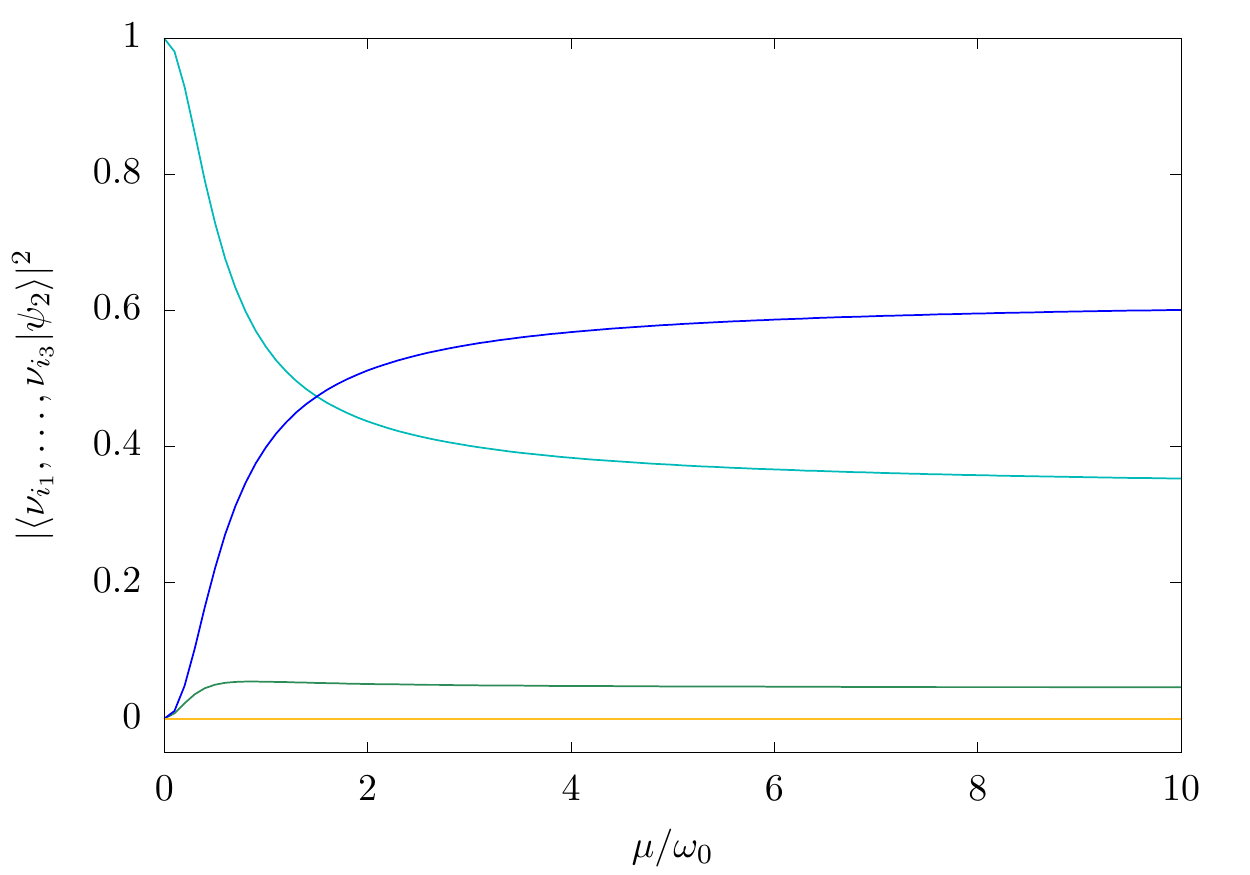}}
	~
	\subfloat[The mass-basis decomposition of the third excited state $\ket{\psi_3}$ in the $N=M=4$ system, as a function of $\mu$. Here $\kappa=2$.\label{N4State}]{
	    \includegraphics[width=0.48\textwidth]{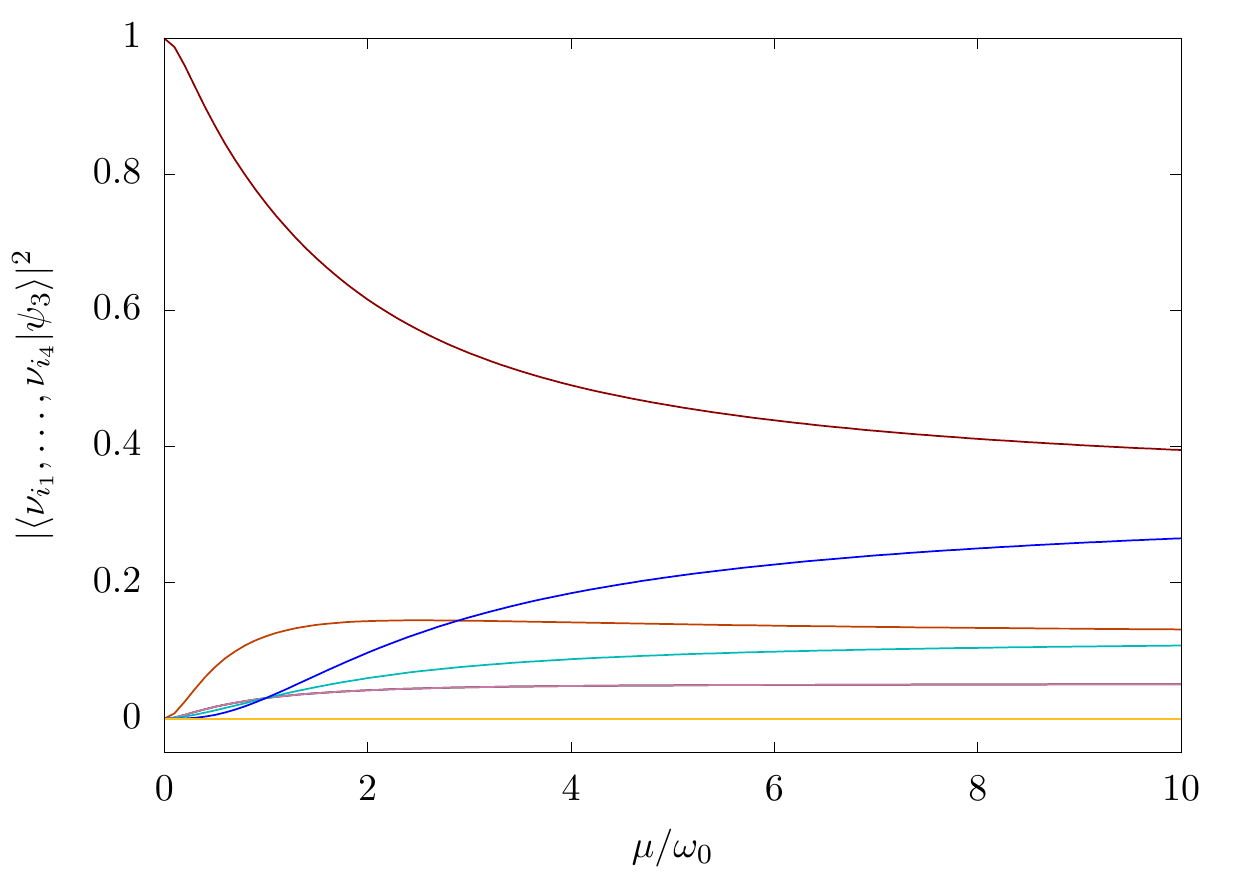}}
\end{center}
	\caption{Overlaps, $|\braket{\nu_{i_1},\ldots,\nu_{i_N}|\psi_n}|^2$, of an excited state $\ket{\psi_n}$ with the mass basis states (with $i_1,\ldots,i_N=1,2$), for systems with different sizes $N$. Here, the $n^\mathrm{th}$ energy eigenstate $\ket{\psi_n}$ ($n=0,\ldots,2^N-1)$ is connected to the mass eigenstate $\ket{\nu_{j_1},\ldots,\nu_{j_N}}$ in the limit $\mu\to0$, where $j_{N+1-k}=1+(k^\mathrm{th}$ digit of $n$ in binary representation$)$. Observe that a state with a given $\kappa$ has ${}^NC_\kappa$ nontrivial components for $\mu>0$. For Figure \ref{N4State}, two of the six nontrivial overlaps are numerically indistinguishable.}
%	\caption{We display the probabilities for excited states with different $N$ of different mass basis states. Here, the $n$th excited energy eigenstate $\ket{\psi}$ with energy $E_n$ ($n=1,\ldots,2^N-1)$ has probabilities $P_n$ of overlap with mass eigenstate $\ket{\nu_{i_1},\ldots,\nu_{i_N}}$ where $i_j=1+(j$th digit of $n$ in binary representation$)$. Observe that a state with a given $\kappa$ has ${}^NC_\kappa$ nontrivial components for $\mu>0$. For Figure \ref{N4State}, two of the six nontrivial probabilities are identical, up to floating point error.}
	\label{States}
\end{figure*}

In addition to our energy eigenvalues, we may calculate energy eigenstates for generic $\mu$ using Eqs. \eqref{eigstdet}--\eqref{coverup}. These eigenstates may be encoded as a sequence of their $2^M$ coefficients in the mass basis. Particular examples of eigenstates for systems with for $M=3,4$ are shown in Fig. \ref{States}. A complete knowledge of the eigenvalues and eigenstates of the system enables calculating the time-evolution (i.e., evolution with $\mu$) of an arbitrary initial state, in the adiabatic limit. This can be used for the purposes of comparing the results of a many-body calculation to the corresponding results in the mean-field limit, as well as for studying exclusive many-body effects like the emergence of quantum entanglement between the various modes as the system evolves.

\section{Conclusions} \label{sec:concl}

Adiabatic evolution of a many-neutrino system in the single-angle approximation is exactly solvable in the sense that such an evolution can be completely characterized by the solutions of the appropriate Bethe ansatz equations. However, solving those nonlinear Bethe ansatz equations is a highly nontrivial problem. In this paper we presented a technique to evaluate the exact adiabatic eigenvalues and eigenstates of the collective neutrino oscillation Hamiltonian by casting Bethe ansatz equations first into a differential form, then into a set of algebraic equations which are numerically more tractable than the original Bethe ansatz equations. With our outlined procedure, to determine solutions for up to $N\sim 15$ for the interesting region of $\mu\leq5$ a personal computer is sufficient. Going to higher values of $N$ would require more computational power. In the future we will explore the behavior of the solutions as $N$ gets larger.

{An immediate benefit of obtaining such exact solutions is the ability to explore the limits of applicability of the commonly used mean-field solutions. Notably, our exact problem (with $j_p=1/2$) has $2^N$ solutions, while the mean-field problem has only $2N$ solutions in total. Clearly, in the mean-field case we are either losing or combining many states.  Earlier studies of the question of entangled neutrinos also explored modifications of the collective oscillations due to the many-body effects present when one goes beyond the one-body description inherent to the mean-field approximation \cite{Bell:2003mg, Friedland:2003dv, Friedland:2003eh}. 
These papers primarily investigate part of the Hamiltonian which survives in the large $\mu$ limit and explore if the oscillations speed up due to many-particle entanglements. The growth rates they obtain differ by a factor of $\sqrt{N}$ relative to each other, depending on the setup. In future publications we plan to explore many-body entanglement effects using our approach, for generic values of $\mu$ where both the one-body and two-body terms in the Hamiltonian play a role. Clearly exploring these issues is a step which needs to be taken before moving on to astrophysics applications.}

Many of the elements heavier than iron were formed by rapidly capturing neutrons on seed nuclei (r-process nucleosynthesis). In the astrophysical site of the r-process nucleosynthesis many reactions take place during a rather short duration, a feature which is typically associated with explosive phenomena. 
Currently the leading candidates for the sites of r-process nucleosynthesis are core-collapse supernovae and binary neutron star mergers. It is known that these nucleosynthesis yields are sensitive to the neutrino flavor evolution inside the supernova envelope. Here we presented an exact procedure to describe neutrino propagation including neutrino-neutrino interactions. It would be also interesting to explore whether any possible differences between the mean-field approximation and the exact adiabatic many-body approach would impact nucleosynthesis yields as well as supernova neutrino detection in terrestrial experiments.

\begin{acknowledgments}
We would like to thank E.~Armstrong, S.~Birol, S.~Coppersmith, G.~Fuller, E.~Grohs, A.  Hashimoto, C. Johnson, E.~Rrapaj, J.~Schmidt, M.~Sen, and S.~Shalgar for valuable conversations. This work was supported in part by the U.S. National Science Foundation Grants PHY-1630782 and PHY-1806368. 
\end{acknowledgments}

\appendix

\section{Deriving the Lambda equations from the Bethe Ansatz} \label{sec:applambda}

In this appendix, we shall present the derivation of the coupled quadratic equations in Lambda, Eq.~\eqref{algebraicLambda}, from the Bethe Ansatz equations Eq.~\eqref{originalBA}. 
These results are dispersed through the condensed-matter physics literature. Here for the convenience of the reader we gather them in one place. 
Using the function $\Lambda(\lambda)$ defined in Eq.~\eqref{Lambda}, one can write
\begin{equation}
\Lambda^2(\lambda) + \Lambda'(\lambda) = \sum_{\substack{\alpha,\beta=1 \\ \alpha\neq \beta}}^\kappa \frac{1}{(\lambda - \zeta_\alpha)(\lambda-\zeta_\beta)}.
\end{equation}

Using partial fraction decomposition and symmetry arguments, this may be rewritten as
\begin{equation}
\begin{split}
\Lambda^2(\lambda) + \Lambda'(\lambda) &= 2\,\sum_{\alpha=1}^\kappa \left[ \frac{1}{\lambda - \zeta_\alpha} \sum_{\beta\neq \alpha}^\kappa \frac{1}{\zeta_\alpha - \zeta_\beta} \right] \\
&\equiv 2\,\sum_{\alpha=1}^\kappa \frac{W(\zeta_\alpha)}{\lambda - \zeta_\alpha}.
\end{split}
\end{equation}

Replacing $W(\zeta_\alpha)$ with the left hand side of the Bethe Ansatz equations Eq.~\eqref{originalBA}, one obtains
\begin{equation}
\Lambda^2(\lambda) + \Lambda'(\lambda) = 2\,\sum_{\alpha=1}^\kappa \frac{1}{\lambda - \zeta_\alpha} \left[ - \frac{1}{2\mu}  - \sum_{p} \frac{j_p}{\omega_p - \zeta_\alpha} \right].
\end{equation}
Using the definition of $\Lambda(\lambda)$ from Eq.~\eqref{Lambda} after changing the order of summation and using partial fraction decomposition, this equation reduces to
\begin{equation} \label{eq:LambdaDEA}
\Lambda^2(\lambda) + \Lambda'(\lambda) +\frac{1}{\mu} \Lambda(\lambda) = 2 \sum_{p} j_p \frac{\Lambda(\lambda) - \Lambda(\omega_p)}{\lambda - \omega_p}.
\end{equation}

It can be shown that the ordinary differential equation, Eq.~(\ref{eq:LambdaDEA}), is exactly equivalent to the Bethe Ansatz equations, i.e., every solution of Eq.~(\ref{eq:LambdaDEA}) corresponds to a unique solution of the Bethe Ansatz equations, and vice versa. This equivalence can be proven using the fact that every step of the above derivation is reversible---one could just as easily start with Eq.~(\ref{eq:LambdaDEA}) and derive the Bethe Ansatz equations, Eq.~(\ref{originalBA}). % It should also be noted that it is not trivial to integrate this ODE, because the right-hand side is not \textit{a priori} determined because of the presence of $\Lambda(\omega_p)$. Finding the solution $\Lambda(\lambda)$ to an ODE with such a structure is a problem that may lend itself to optimization-based techniques.

%\subsection{From the ODE to a set of algebraic equations in $\Lambda(\omega_p)$} \label{sec:Lambdaeq}
\vskip 0.5cm

The ordinary differential equation in Eq.~(\ref{eq:LambdaDEA}) may be converted to a set of algebraic equations in $\Lambda(\omega_p)$, for each $\omega_p$ in the system. In order to do so, one can Taylor-expand $\Lambda(\lambda)$ around $\lambda = \omega_q$. Following this Taylor expansion, the 
right-hand side of Eq.~(\ref{eq:LambdaDEA}) may be written as 
\begin{equation}
\begin{split}
	2 \sum_{p} j_p \frac{\Lambda(\lambda) - \Lambda(\omega_p)}{\lambda - \omega_p} = 2 \sum_{p \neq q} j_p \frac{\Lambda(\lambda) - \Lambda(\omega_p)}{\lambda - \omega_p} \\
	+  2 j_q \left[\Lambda'(\omega_q) + \frac{1}{2} \Lambda''(\omega_q) (\lambda - \omega_q) + \ldots \right]
\end{split}
\label{eq:Lambdataylor}
\end{equation}
where $\Lambda'(\omega_q), \Lambda''(\omega_q), \ldots$ are the successive derivatives of $\Lambda(\lambda)$ with respect to $\lambda$, as evaluated at $\lambda = \omega_q$. Using this expansion and then taking the limit $\lambda \rightarrow \omega_q$, one obtains the following set of equations (one for each $\omega_q$)
\begin{equation} \label{eq:Lambdaeq}
\begin{split}
\Lambda^2(\omega_q) + (1- 2j_q) \Lambda'(\omega_q) +\frac{1}{\mu} \Lambda(\omega_q) \\ 
= 2 \sum_{p \neq q} j_p \frac{\Lambda(\omega_q) - \Lambda(\omega_p)}{\omega_q - \omega_p}.
\end{split}
\end{equation}

If $j_q = 1/2$ for each $q$ (corresponding to there being just one neutrino per bin), then the $\Lambda'$ term vanishes and the equations become purely algebraic (Eq.~\eqref{algebraicLambda}), and can in principle be solved to obtain the solutions $\{\Lambda(\omega_q): q=1,\ldots,M \}$. If $j_q > 1/2$, then one can take successive derivatives of Eq.~(\ref{eq:LambdaDEA}) with respect to $\lambda$, as noted in Refs.~\cite{Babelon_2007,Faribault:2011rv,Araby:2012ru,Claeys:2015}. For example, taking the first derivative of Eq.~(\ref{eq:LambdaDEA}) followed by the $\lambda \rightarrow \omega_q$ limit gives us
\begin{equation} \label{eq:Lambdapeq}
\begin{split}
2 \Lambda(\omega_q) \Lambda'(\omega_q) + (1- j_q) \Lambda''(\omega_q) + \frac{1}{\mu} \Lambda'(\omega_q) \\
= 2 \sum_{p \neq q} j_p \left[ \frac{\Lambda'(\omega_q)}{\lambda - \omega_p} - \frac{\Lambda(\omega_q) - \Lambda(\omega_p)}{(\omega_q - \omega_p)^2} \right].
\end{split}
\end{equation}

Now, if $j_q = 1$, then the $\Lambda''$ term vanishes, and one may use Eqs.~(\ref{eq:Lambdaeq}) and (\ref{eq:Lambdapeq}) to eliminate $\Lambda'(\omega_q)$ from the system of equations. Likewise, if $j_q = 3/2$, one may obtain an additional equation by taking the \textit{second} derivative of Eq.~(\ref{eq:LambdaDEA}), and use it along with Eqs.~(\ref{eq:Lambdaeq}) and (\ref{eq:Lambdapeq}) to eliminate $\Lambda'(\omega_q)$ and $\Lambda''(\omega_q)$. And so on, for higher values of $j_q$.

\section{Proof of the applicability of the homotopy continuation method} \label{sec:apphomotopy}

To carry out this homotopy method, let us first rewrite Eq.~\eqref{tildeLambda} as the system $\vec{F}(\vec{z})=\vec{0}$ using the functions 
\begin{equation}
F_p(\vec{z})=z_p^2+z_p-\mu\sum_{q\neq p}\frac{z_p-z_q}{\eta_{pq}}
\label{functionF}
\end{equation}
for $p=1,\ldots,M$. In particular, since we know each value of $\widetilde{\Lambda}_p$ must be real\footnote{Since the eigenvalues of a Hermitian operator are real (cf. Eqs. (\ref{LambdaX}) and (\ref{eigenvaluesLambda})), any complex-valued solutions $\zeta_\alpha$ of Eq.~\eqref{originalBA} must come in complex conjugate pairs~\cite{Pehlivan:2011hp,Birol:2018qhx}, implying by Eq.~\eqref{Lambda} that $\Lambda(\omega_p)$ and therefore $\widetilde{\Lambda}(\omega_p)$ must be real.}
, we restrict $\vec{z}\in\mathbb{R}$. Additionally, we can define a set of functions $\vec{G}$ with simpler solutions by
\begin{equation}
G_p(\vec{z})=z_p^2+z_p
\label{functionG}
\end{equation}
and the family of functions $\vec{\mathcal{F}}(\vec{z},t)=(1-t)\vec{G}(\vec{z})+t\vec{F}(\vec{z})$ for $0\leq t\leq1$. Then, from the explicit form of 
\begin{equation}
\mathcal{F}_p(\vec{z},t)=z_p^2+z_p-t\mu\sum_{q\neq p}\frac{z_p-z_q}{\eta_{pq}}
\label{functionH}
\end{equation}
we see that $\vec{\mathcal{F}}$ is an analytic (and therefore a $C^2$) mapping of $\mathbb{R}^M\times[0,1]\to\mathbb{R}^M$. By Sard's Theorem, we find that $\vec{\mathcal{F}}$ is a regular function. Recall that regular, in the context of algebraic geometry, means that: (i) $\vec{\mathcal{F}}(\vec{z}_0,t)=\vec{0}$ implies $\partial_{\vec{z}}\vec{\mathcal{F}}(\vec{z}_0,t)$ has rank $M$ for fixed $t=0,1$ at almost all $\vec{z}_0\in\mathbb{R}^M$ and (ii) $\vec{\mathcal{F}}(\vec{z_0},t)=\vec{0}$ implies $\partial_{\vec{z},t}\vec{\mathcal{F}}(\vec{z_0},t)$ has rank $M$ for fixed $0<t<1$ at almost all $\vec{z}_0\in\mathbb{R}^M$. (To clarify, $\partial_{\vec{z},t}\vec{\mathcal F}$ is the total Jacobian of $\vec{H}$ while $\partial_{\vec{z}}\vec{\mathcal F}$ is the Jacobian of $\vec{\mathcal F}$ stripped of its final column, $\partial_t\vec{\mathcal F}$.) Further, a theorem by Garcia and Zangwill \cite{Garcia1979} states that if, in addition, for all sequences with $||\vec{z}^{(n)}||\to\infty$ there is a $p$ such that $F_p,G_p$ satisfy
\begin{equation}
\lim_{m\to\infty}\frac{F_p(\vec{z}^{(n_m)})}{G_p(\vec{z}^{(n_m)})}\nless\alpha
\label{GarciaCondn}
\end{equation}
for some $\alpha>0$ and a sub-sequence of points $\vec{z}^{(n_m)}$ at which $G_p(\vec{z}^{(n_m)})\neq0$, then all roots of $\vec{F}$ can be found one-to-one from the known roots of $\vec{G}$ (i.e., the asymptotic solutions as $\mu\to0$) by homotopy curves as we vary $t=0$ to $t=1$. The symbol $\nless$ implies that the real part of the limit (which generically may be complex if our chosen $\vec{F}$ were not strictly real) must be greater than or equal to a positive $\alpha$ that we are free to choose. We can see that this additional hypothesis is true for our system by first calculating for an arbitrary $p$ that
\begin{equation}
\frac{F_p(\vec{z}^{(n)})}{G_p(\vec{z}^{(n)})}=1-\frac{\mu\sum_{q\neq p}(z_p^{(n)}-z_q^{(n)})/\eta_{pq}}{z_p^{(n)2}+z_p^{(n)}}
\label{GarciaCondnCheck}
\end{equation}
for any sequence of points $\vec{z}^{(n)}$ after eliminating any of the points in the sequence solving $G_p(\vec{z})=0$ for some $p$. (We know we may choose such a $p$ since we know there are finitely many points solving $G_p(\vec{z})=0$ for all $p$.) Moreover, we may observe that this sequence must have at least one $r$ for which $|z_r^{(n)}|\to\infty$ at the fastest rate amongst the $\vec{z}^{(n)}$. Selecting $p=r$ we find that $F_p(\vec{z}^{(n)})/G_p(\vec{z}^{(n)})\to1$ on this sub-sequence. Thus, we have shown our desired system $\vec{F}$ is homotopic to $\vec{G}$ as we vary $t\mu$ from $0$ to $\mu$, in the sense described by Ref.~\cite{Garcia1979}, and so all our solutions to $\vec{F}(\vec{z})=\vec{0}$ can be found from those of $\vec{G}(\vec{z})=\vec{0}$ via a single homotopy parameter. Thus we justify our use of the algebraic homotopy method for finding our $\widetilde{\Lambda}(\omega_p)$ solutions.

Additionally, we confirm from this process that there must be exactly $2^M$ solutions to Eq.~\eqref{tildeLambda} for $\mu\neq0$ as well. Moreover, for a given $\kappa=-\sum_{p=1}^M\widetilde{\Lambda}_p$, there are $^MC_\kappa$ solutions. We can observe that there are equally as many ways for us to permute the values $0,-1$ amongst our different $\widetilde{\Lambda}_p$ for a given $\kappa$. Thus, we expect that by considering all possible permutations of $0,-1$ for each $\widetilde{\Lambda}_p$ at $\mu=0$ we may find all $\sum_{\kappa=0}^M{} ^MC_\kappa =2^M$ solutions.

Lastly, consider the situation in which we fix the values of each $\omega_p$ while we allow $\mu\geq0$ to increase. Observe from our above discussion of the homotopy method that the maximum value of $\mu$ (i.e., $t\mu$ with $t=1$) in Eq.~\eqref{functionH} is simply an arbitrary real $\mu>0$. So, we expect that we may obtain not only the numerical solutions to $\vec{F}(\vec{z})=\vec{\mathcal F}(\vec{z},1)=\vec{0}$ but also the $n^\mathrm{th}$ ($n=0,1\ldots$) intermediate numerical solutions $\vec{z}^{(n)}$ to $\vec{\mathcal F}(\vec{z}^{(n)},t^{(n)})=\vec{0}$ for each $0\leq t^{(n)}\leq 1$, provided that our steps $t^{(n+1)}-t^{(n)}>0$ are sufficiently small for the discrete steps of our homotopy method to be appropriate. In fact, by applying this method to only $\vec{F}|_{\mu=\mu_{\max}}$, we may obtain the solutions to Eq.~\eqref{tildeLambda} for all $0\leq\mu\leq\mu_{\max}$ with an arbitrary maximum $\mu_{\max}$, as $\vec{z}^{(n)}$ may be interpreted as the solutions to $\vec{F}|_{\mu=t^{(n)}\mu_{\max}}$.

\section{Derivations of Eigenstates for \texorpdfstring{$\kappa=2,3$}{kappa=2,3}} \label{sec:appeig23}

We can demonstrate the claim from Section ~\ref{sec:eiglambda} that our general eigenstates can be written in terms of $\Lambda_p$ in lieu of $\zeta_\alpha$ for $\kappa=2,3$ when $j_p=1/2$ for all $p$. Here we present two ways to derive the eigenstate expressions for $\kappa=2,3$. First note that, for all $\kappa$, we have the identity: 
\begin{eqnarray}
\frac{1}{\kappa!}\bigg(\sum_{\alpha=1}^\kappa S^-_\alpha\bigg)^\kappa\ket{j,+j} &=& \frac{1}{\kappa!}\bigg(\sum_{p=1}^M \Lambda_pJ_{p}^-\bigg)^\kappa\ket{j,+j} 
\nonumber \\
 	 = e_\kappa \big(\Lambda_1J_{1}^-,&\ldots&,\Lambda_MJ_{M}^-\big)\ket{j,+j}
\label{fullycrossedterm}
\end{eqnarray}
where $e_\kappa$ is the degree-$\kappa$ elementary symmetric polynomial in $M$ variables. Now, for $\kappa=2$ we have
\begin{align}
&\sum_{\alpha=1}^2(S^-_\alpha)^2\ket{j,+j}\nonumber \\ &= \sum_{\alpha=1}^2 \bigg(\sum_{p=1}^M\frac{J_p^-}{\omega_p-\zeta_\alpha}\bigg)\bigg(\sum_{p=1}^M\frac{J_p^-}{\omega_p-\zeta_\alpha}\bigg)\ket{j,+j} 
\nonumber \\
&= \sum_{\alpha=1}^2\sum_{p=1}^M\sum_{\substack{q=1\\q\neq p}}^M \frac{1}{(\omega_p-\zeta_\alpha)(\omega_1-\zeta_\alpha)}J_p^-J_q^-\ket{j,+j} 
\nonumber \\
&=  \sum_{p=1}^M\sum_{\substack{q=1\\q\neq p}}^M\sum_{\alpha=1}^2\bigg(\frac{1}{\omega_p-\zeta_\alpha}-\frac{1}{\omega_q-\zeta_\alpha}\bigg)\frac{-1}{\omega_p-\omega_q}J_p^-J_q^-\ket{j,+j} 
\nonumber \\
&= -\sum_{\substack{p,q=1\\q\neq p}}^M\frac{\Lambda_p-\Lambda_q}{\omega_p-\omega_q}J_p^-J_q^-\ket{j,+j} 
\label{noncrossedterm2}
\end{align}
and so
\begin{widetext}
\begin{equation}
S^-_1S^-_2\ket{j,+j} = \frac{1}{2}\bigg[\bigg(\sum_{\alpha=1}^2S^-_\alpha\bigg)^2-\sum_{\alpha=1}^2(S^-_\alpha)^2\bigg]\ket{j,+j} 
= \frac{1}{2}\sum_{\substack{p,q=1\\p\neq q}}^M\bigg(\Lambda_p\Lambda_q+\frac{\Lambda_p-\Lambda_q}{\omega_p-\omega_q}\bigg)J_p^-J_q^-\ket{j,+j} .
\label{kappa2eigenstates}
\end{equation}
\end{widetext}

Thus, we have rewritten the eigenstates as well as the eigenvalues of the Hamiltonian in Eq. \eqref{eq:saham} completely in terms of our defined $\Lambda(\omega_p)$ parameters, bypassing the need to directly compute the Bethe Ansatz variables $\zeta_\alpha$. This process can be carried out for $\kappa=3$ similarly:
\begin{widetext}
\begin{eqnarray}
&&\sum_{\alpha=1}^3(S^-_\alpha)^3\ket{j,+j} =  \sum_{\alpha=1}^3\sum_{\substack{p,q,r=1\\p,q,r\mathrm{\,distinct}}}^M  \frac{J_p^-J_q^-J_r^-}{(\omega_p-\zeta_\alpha)(\omega_q-\zeta_\alpha)(\omega_r-\zeta_\alpha)}\ket{j,+j} %q\neq p,r\neq p,q
\nonumber \\
&=& \sum_{\substack{p,q,r=1\\p,q,r\mathrm{\,distinct}}}^M \sum_{\alpha=1}^3  \bigg[\frac{-1}{\omega_p-\omega_q}\bigg(\frac{1}{\omega_p-\zeta_\alpha} - \frac{1}{\omega_q-\zeta_\alpha}\bigg)\frac{1}{\omega_r-\zeta_\alpha}\bigg]J_p^-J_q^-J_r^-\ket{j,+j}
\nonumber \\
\hspace*{-1cm}
&=& \sum_{\substack{p,q,r=1\\p,q,r\mathrm{\,distinct}}}^M  \frac{-1}{\omega_p-\omega_q} \sum_{\alpha=1}^3\bigg[\frac{-1}{\omega_p-\omega_r}\bigg(\frac{1}{\omega_p-\zeta_\alpha}-\frac{1}{\omega_r-\zeta_\alpha}\bigg)
%\nonumber \\
%\phantom{\sum_{\substack{p,q,r=1\\p,q,r\mathrm{\,distinct}}}^M\frac{1}{\omega_p-\omega_q} %\sum_{\alpha=1}^3  \bigg[\frac{-1}{\omega_p-\omega_r}\bigg(\frac{1}{\omega_p-\zeta_\alpha}-}
-\frac{-1}{\omega_q-\omega_r}\bigg(\frac{1}{\omega_q-\zeta_\alpha}-\frac{1}{\omega_r-\zeta_\alpha}\bigg)\bigg] \times J_p^-J_q^-J_r^-\ket{j,+j}
\nonumber \\
&=& \sum_{\substack{p,q,r=1\\p,q,r\mathrm{\,distinct}}}^M\frac{1}{\omega_p-\omega_q} \bigg(\frac{\Lambda_p-\Lambda_r}{\omega_p-\omega_r}-\frac{\Lambda_q-\Lambda_r}{\omega_q-\omega_r}\bigg)J_p^-J_q^-J_r^-\ket{j,+j}
\end{eqnarray}
and
\begin{align}
\sum_{\alpha=1}^3(S^-_\alpha)^2\sum_{\beta\neq \alpha}S^-_\beta\ket{j,+j} &= \sum_{\alpha=1}^3\sum_{\substack{p,q,r=1\\p,q,r\mathrm{\,distinct}}}^M\frac{J_p^-J_q^-}{(\omega_p-\zeta_\beta)(\omega_q-\zeta_\gamma)}\bigg(\frac{J_r^-}{\omega_r-\zeta_k}-\frac{J_r^-}{\omega_r-\zeta_l}\bigg)\ket{j,+j} 
\nonumber \\
&= \sum_{\substack{p,q,r=1\\p,q,r\mathrm{\,distinct}}}^M \sum_{i=1}^3\frac{-1}{\omega_p-\omega_q}\bigg(\frac{1}{\omega_p-\zeta_\alpha}-\frac{1}{\omega_q-\zeta_\alpha}\bigg)\bigg[\Lambda_r-\frac{1}{\omega_r-\zeta_\alpha}\bigg] J_p^-J_q^-J_r^-\ket{j,+j}
\nonumber \\
&= -\sum_{\substack{p,q,r=1\\p,q,r\mathrm{\,distinct}}}^M \bigg[\Lambda_r\frac{\Lambda_p-\Lambda_q}{\omega_p-\omega_q}-\frac{1}{\omega_p-\omega_q}\sum_{\alpha=1}^3\bigg(\frac{1}{\omega_p-\zeta_\alpha}-\frac{1}{\omega_q-\zeta_\alpha}\bigg)\frac{1}{\omega_r-\zeta_\alpha}\bigg] J_p^-J_q^-J_r^-\ket{j,+j} 
\nonumber \\
&= -\sum_{\substack{p,q,r=1\\p,q,r\mathrm{\,distinct}}}^M\bigg[\Lambda_r\frac{\Lambda_p-\Lambda_q}{\omega_p-\omega_q}+\frac{1}{\omega_p-\omega_q}\bigg(\frac{\Lambda_p-\Lambda_r}{\omega_p-\omega_r}-\frac{\Lambda_q-\Lambda_r}{\omega_q-\omega_r}\bigg)\bigg] J_p^-J_q^-J_r^-\ket{j,+j} 
\end{align}
where in the first line we are using $\beta\neq \gamma$ and $\beta,\gamma\neq \alpha$. Thus, 
\begin{align}
S_1^-S_2^-S_3^-\ket{j,+j} &= \frac{1}{3!}\bigg[\bigg(\sum_{\alpha=1}^3 S_\alpha^-\bigg)^3-\sum_{\alpha=1}^3(S_\alpha^-)^3-3\sum_{\alpha=1}^3(S_\alpha^-)^2\sum_{\beta\neq \alpha}S_\beta^-\bigg]\ket{j,+j} 
\nonumber \\
&= \frac{1}{3!}\sum_{\substack{p,q,r=1\\p,q,r\mathrm{\,distinct}}}^M\bigg[\Lambda_p\Lambda_q\Lambda_r+3\Lambda_r\frac{\Lambda_p-\Lambda_q}{\omega_p-\omega_q} +\frac{2}{\omega_p-\omega_q}\bigg(\frac{\Lambda_p-\Lambda_r}{\omega_p-\omega_r}-\frac{\Lambda_q-\Lambda_r}{\omega_q-\omega_r}\bigg)\bigg]J_p^-J_q^-J_r^-\ket{j,+j} .
\label{kappa3eigenstates}
\end{align}
There is a more direct approach to arrive at the same results, by decomposing terms of $S_1^-S_2^-$ and $S_1^-S_2^-S_3^-$ immediately without consideration of other products -- as we will do in the next section. 
Also, note that the form of these results bears similarity to the characters of exterior powers of vector spaces (i.e., $\Lambda^\kappa V$ for $\kappa=2,3$), suggesting a possibility that knowledge of these characters may allow us to rewrite $\prod_{\alpha=1}^\kappa S_\alpha^-$ in terms of $\Lambda_p$ more swiftly.

\section{Alternative Derivation of Eigenstates for \texorpdfstring{$\kappa=2,3$}{kappa=2,3}} \label{sec:appeig23_2}

Here, we present a more direct method of evaluating the product $\prod_{i=1}^\kappa S_\alpha^-\ket{j,+j}$ for the cases of $\kappa=2,3$. In $\kappa=2$, we can quickly compute:
\begin{align}
S_1^-S_2^-\ket{j,+j} &= \frac{1}{2}(S_1^-S_2^-+S_2^-S_1^-)\ket{j,+j}
\nonumber \\
	&= \frac{1}{2}\sum_{p,q=1}^M\bigg[\frac{1}{(\omega_p-\zeta_1)(\omega_q-\zeta_2)}+\frac{1}{(\omega_p-\zeta_2)(\omega_q-\zeta_1)}\bigg]J_p^-J_q^-\ket{j,+j}
\nonumber \\
	&= \frac{1}{2}\sum_{\substack{p,q=1\\p\neq q}}^M\bigg[\frac{1}{(\omega_p-\zeta_1)}\bigg(\Lambda_q-\frac{1}{\omega_q-\zeta_1}\bigg)+\frac{1}{(\omega_p-\zeta_2)}\bigg(\Lambda_q-\frac{1}{\omega_q-\zeta_2}\bigg)\bigg]J_p^-J_q^-\ket{j,+j}
\nonumber \\
	&= \frac{1}{2}\sum_{\substack{p,q=1\\p\neq q}}^M\bigg(\Lambda_p\Lambda_q+\frac{\Lambda_p-\Lambda_q}{\omega_p-\omega_q}\bigg)J_p^-J_q^-\ket{j,+j} .
	\label{kappa2eigenstatesA}
\end{align}

Analogously, we can extend this argument to $\kappa=3$:
\begin{align}
S_1^-S_2^-S_3^-\ket{j,+j} &= \sum_{\substack{p,q,r=1\\p,q,r\mathrm{\,distinct}}}^M\frac{J_p^-J_q^-J_r^-}{(\omega_p-\zeta_1)(\omega_q-\zeta_2)(\omega_r-\zeta_3)}\ket{j,+j}
\nonumber \\
	&= \sum_{\substack{p,q,r=1\\p,q,r\mathrm{\,distinct}}}^M\frac{1}{(\omega_p-\zeta_1)(\omega_q-\zeta_2)}\bigg(\Lambda_r-\frac{1}{\omega_r-\zeta_1}-\frac{1}{\omega_r-\zeta_2}\bigg)J_p^-J_q^-J_r^-\ket{j,+j}
\nonumber \\
&= \frac{1}{3!}\sum_{\sigma\in\mathrm{Sym}(3)}\sum_{\substack{p,q,r=1\\p,q,r\mathrm{\,distinct}}}^M\frac{1}{(\omega_p-\zeta_{\sigma(1)})(\omega_q-\zeta_{\sigma(2)})}\bigg(\Lambda_r-\frac{1}{\omega_r-\zeta_{\sigma(1)}}-\frac{1}{\omega_r-\zeta_{\sigma(2)}}\bigg)J_p^-J_q^-J_r^-\ket{j,+j}
\nonumber \\
\label{kappa3eigenstatesA}
\end{align}
while 
\begin{align}
\sum_{\substack{p,q,r=1\\p,q,r\mathrm{\,distinct}}}^M\frac{\Lambda_r}{(\omega_p-\zeta_1)(\omega_q-\zeta_2)}J_p^-J_q^-J_r^-\ket{j,+j} &\to \frac{1}{3}\sum_{\substack{p,q,r=1\\p,q,r\mathrm{\,distinct}}}^M\Lambda_r\bigg[\frac{1}{(\omega_p-\zeta_1)(\omega_q-\zeta_2)}+\frac{1}{(\omega_p-\zeta_2)(\omega_q-\zeta_3)}
\nonumber \\
	&\phantom{\frac{1}{3}\sum_{\substack{p,q,r=1\\p,q,r\mathrm{\,distinct}}}^M\Lambda_r\bigg[\frac{1}{(\omega_p-\zeta_1)}} +\frac{1}{(\omega_p-\zeta_3)(\omega_q-\zeta_1)}\bigg]J_p^-J_q^-J_r^-\ket{j,+j}
\nonumber \\
\implies \sum_{\substack{p,q,r=1\\p,q,r\mathrm{\,distinct}}}^M\frac{\Lambda_r}{(\omega_p-\zeta_1)(\omega_q-\zeta_2)}J_p^-J_q^-J_r^-\ket{j,+j} &\to \frac{1}{2}\sum_{\substack{p,q,r=1\\p,q,r\mathrm{\,distinct}}}^M\frac{\Lambda_r}{\omega_p-\zeta_3}\bigg(\Lambda_q-\frac{1}{\omega_q-\zeta_3}\bigg)J_p^-J_q^-J_r^-\ket{j,+j}
\nonumber
\end{align}
\begin{align}
\implies \sum_{\sigma\in\mathrm{Sym}(3)}\sum_{\substack{p,q,r=1\\p,q,r\mathrm{\,distinct}}}^M\frac{\Lambda_r}{(\omega_p-\zeta_{\sigma(1)})(\omega_q-\zeta_{\sigma(2)})}J_p^-J_q^-J_r^-\ket{j,+j} &= \sum_{\substack{p,q,r=1\\p,q,r\mathrm{\,distinct}}}^M\Lambda_r\bigg(\Lambda_p\Lambda_q+\frac{\Lambda_p-\Lambda_q}{\omega_p-\omega_q}\bigg)J_p^-J_q^-J_r^-\ket{j,+j}
\end{align}
and 
\begin{align}
&\sum_{\substack{p,q,r=1\\p,q,r\mathrm{\,distinct}}}^M\frac{1}{(\omega_p-\zeta_1)(\omega_q-\zeta_2)(\omega_r-\zeta_1)}J_p^-J_q^-J_r^-\ket{j,+j} \to \frac{1}{12}\sum_{\sigma\in\mathrm{Sym}(3)}\sum_{\substack{p,q,r=1\\p,q,r\mathrm{\,distinct}}}^M\frac{1}{(\omega_p-\zeta_{\sigma(1)})(\omega_r-\zeta_{\sigma(1)})}
\nonumber \\
	&\phantom{\to \frac{1}{12}\sum_{\sigma\in\mathrm{Sym}(3)}} \times\bigg(\frac{1}{\omega_q-\zeta_{\sigma(2)}}+\frac{1}{\omega_q-\zeta_{\sigma(3)}}\bigg)J_p^-J_q^-J_r^-\ket{j,+j}
\nonumber \\
	&= \frac{1}{12}\sum_{\sigma\in\mathrm{Sym}(3)}\sum_{\substack{p,q,r=1\\p,q,r\mathrm{\,distinct}}}^M\frac{1}{(\omega_p-\zeta_{\sigma(1)})(\omega_r-\zeta_{\sigma(1)})}
%\nonumber \\
%	\phantom{= \frac{1}{12}\sum_{\sigma\in\mathrm{Sym}(3)}}
\times\bigg(\Lambda_q-\frac{1}{\omega_q-\zeta_{\sigma(1)}}\bigg)J_p^-J_q^-J_r^-\ket{j,+j}
\nonumber \\
	&= \frac{1}{6}\sum_{\substack{p,q,r=1\\p,q,r\mathrm{\,distinct}}}^M\bigg(\Lambda_q\frac{\Lambda_p-\Lambda_r}{\omega_p-\omega_r}+\frac{\frac{\Lambda_p-\Lambda_q}{\omega_p-\omega_r}-\frac{\Lambda_q-\Lambda_q}{\omega_q-\omega_r}}{\omega_p-\omega_r}\bigg)J_p^-J_q^-J_r^-\ket{j,+j} .
\end{align}

Thus, we arrive at
\begin{align}
S_1^-S_2^-S_3^-\ket{j,j} &= \frac{1}{3!}\sum_{\substack{p,q,r=1\\p,q,r\mathrm{\,distinct}}}\bigg[\Lambda_p\Lambda_q\Lambda_r+3\Lambda_r\frac{\Lambda_p-\Lambda_q}{\omega_p-\omega_q} +\frac{2}{\omega_p-\omega_q}\bigg(\frac{\Lambda_p-\Lambda_r}{\omega_p-\omega_r}-\frac{\Lambda_q-\Lambda_r}{\omega_q-\omega_r}\bigg)\bigg]J_p^-J_q^-J_r^-\ket{j,j} .
\label{kappa3eigenstatesF}
\end{align}
\end{widetext}

\bibliography{notes1}%,allref}

\end{document}